\documentclass[english,showpacs,superscriptaddress,nofootinbib,aps,notitlepage]{revtex4-1}
\usepackage[T1]{fontenc}
\usepackage[latin9]{inputenc}
\usepackage{color}
\usepackage{babel}
\usepackage{amsmath}
\usepackage{amssymb}
\usepackage{stmaryrd}
\usepackage{graphicx}
\usepackage{setspace}
\usepackage[unicode=true,pdfusetitle,
 bookmarks=true,bookmarksnumbered=false,bookmarksopen=false,
 breaklinks=false,pdfborder={0 0 1},backref=false,colorlinks=true]
 {hyperref}

\makeatletter
\usepackage{babel}

\makeatother

\begin{document}
\title{Anisotropic flows of identified hadrons in the equal-velocity quark
combination model at RHIC energy}
\author{Yan-ting Feng}
\affiliation{School of Physics and Physical Engineering, Qufu Normal University,
Shandong 273165, China}
\author{Rui-qin Wang}
\affiliation{School of Physics and Physical Engineering, Qufu Normal University,
Shandong 273165, China}
\author{Feng-lan Shao}
\email{shaofl@mail.sdu.edu.cn}

\affiliation{School of Physics and Physical Engineering, Qufu Normal University,
Shandong 273165, China}
\author{Jun Song }
\email{songjun2011@jnxy.edu.cn}

\affiliation{School of Physics and Electronic Engineering, Jining University, Shandong
273155, China}
\begin{abstract}
We employ an equal-velocity quark combination model to study anisotropic
flows $v_{2}$, $v_{3}$ and $v_{4}$ of identified hadrons at mid-rapidity
in heavy-ion collisions at RHIC energies. Under the equal-velocity
combination mechanism of constituent quarks at hadronization, we build
analytical formulas of anisotropic flows of hadrons in terms of those
of quarks just before hadronization. We systematically analyze the
contribution of higher order flows of quarks, and show how simple
formulas of $v_{2}$, $v_{3}$ and $v_{4}$ of identified hadrons
with the desired precision can be obtained by neglecting the small
contribution of higher order flows of quarks. We systematically test
these simple formulas of hadronic flows by the experimental data of
$v_{2}$, $v_{3}$ and $v_{4}$ of identified hadrons $\phi$, $\Lambda$,
$\Xi^{-}$, $\Omega^{-}$, $\bar{\Lambda}$, $\bar{\Xi}^{+}$, $\bar{\Omega}^{+}$,
$p$ and $\bar{p}$ in Au+Au collisions at $\sqrt{s_{NN}}=$ 19.6,
54.4 and 200 GeV, and we find that the equal-velocity quark combination
model can well describe the measured $v_{2}$, $v_{3}$ and $v_{4}$
of identified hadrons in Au+Au collisions at those collision energies.
We further study the obtained anisotropic flows of quarks and find
two scaling properties which can be qualitatively
understood by the hydrodynamic evolution of thermal quark medium produced
in relativistic heavy-ion collisions. 
\end{abstract}
\maketitle

\section{Introduction\label{sec:Intro} }

Anisotropic transverse flows of produced particles are important physical observables in non-central heavy-ion collisions and were proposed in early years to detect the property of early evolution of thermal partonic medium in relativistic heavy-ion collisions \citep{Ollitrault:1992bk,Voloshin:1994mz,Sorge:1996pc,Poskanzer:1998yz}.  Now, anisotropic transverse flows of hadrons and leptons are widely measured in collisions of different nuclei at different collision energies at colliders such as SPS, RHIC and LHC \citep{Klochkov:2018xvw,Kashirin:2020nvw,STAR:2004jwm,Dixit:2022geb,STAR:2013eve,ALICE:2011ab}, and are widely studied in various theoretical models and/or event generators \citep{Bleicher:2000sx,Lin:2001zk,Lin:2004en,Molnar:2003ff,Gale:2013da,Rode:2019pey,Zhao:2021vmu}.  For example, the hydrodynamic models focus on the evolution of transverse flows and their response to the initial collision geometry and the equation of state of the hot medium produced in collisions \citep{Gale:2013da,Alqahtani:2017mhy}. The theoretical calculations of hydrodynamic models are mainly of the particle flow at the low transverse momentum ($p_{T}$). Microscopic transport models such as UrQMD and AMPT \citep{Tu:2018ora,OrjuelaKoop:2015jss} can further model the effect of rescatterings of hadrons in the late stage of collisions on the anisotropic flow of hadrons. 

In the intermediate $p_{T}$ region, it is found in experiments that
the anisotropic flows of hadrons, in particular the elliptic flows
of hadrons, exhibit a number of constituent quark (NCQ) scaling property
\citep{STAR:2005npq,PHENIX:2006dpn,STAR:2013ayu}, which are generally
regarded to be related to the coalescence/(re-)combination mechanism
of the thermal/soft quarks at hadronization \citep{Kolb:2004gi,Molnar:2003ff,Fries:2003kq,Zhao:2020wcd}.
In most experimental and theoretical studies on the NCQ of hadronic
flows \citep{STAR:2013ayu,STAR:2010ico,Voloshin:2002wa}, the kinetic
variable $p_{T}$ or the transverse kinetic energy $m_{T}-m$ is divided
by the number of constituent (anti-)quarks contained in hadron, which
does not distinguish the possible difference in momentum carried by
quarks of different flavor. In previous work \citep{Song:2020uvu},
we suggest a different description of $v_{2}$ of hadrons and apply
an equal-velocity quark combination model to calculate the elliptic
flow ($v_{2}$) of identified hadrons in relativistic heavy-ion collisions.
Because the momentum of quark is proportional to its constituent mass
in equal-velocity combination (EVC) mechanism, transverse momentum
of up/down quarks and that of strange quarks are explicitly different
in combination and also in the formed hadron. We find that $v_{2}$
of $\phi$ and various baryons in Au+Au collisions at several RHIC
energies and Pb-Pb collisions at $\sqrt{s_{NN}}=$ 2.76 and 5.02 TeV
can be self-consistently understood in terms of the flows of up/down
quarks and that of strange quarks under the EVC mechanism. 

Recently, the STAR collaboration published the precise measurement of
$v_{2}$, $v_{3}$ and $v_{4}$ of identified hadrons in Au+Au collisions
at the intermediate RHIC energies \citep{Dixit:2022geb,STAR:2015rxv,STAR:2022tfp,Sharma:2023dkg,STAR:2022ncy,STAR:2015gge,STAR:2013ayu,Sharma:2015gqe,Li:2010hea},
which can provide the deep insights into mechanism of hadron production
and the property of thermal quark medium produced at these collision
energies. In this paper, we apply an equal-velocity quark combination
model to systematically study $v_{2}$, $v_{3}$ and $v_{4}$ of identified
hadrons $\phi$, $p$, $\Lambda$, $\Xi$, $\Omega$ in Au+Au collisions
at $\sqrt{s_{NN}}=$ 19.6, 54.4 and 200 GeV with particular emphasis
on the physical self-consistency of the mechanism in explaining the experimental data of $v_{2}$, $v_{3}$ and $v_{4}$
of different hadrons measured by the STAR collaboration. In particular, we derive
the analytic formulas of flows of hadrons as functions of those
of quarks just before hadronization and demonstrate how can we obtain
the simple formulas of $v_{2}$, $v_{3}$ and $v_{4}$ of hadrons
with the desired precision by properly analysis of the contribution
of high order flows of quarks. We systematically test these simple
formulas of $v_{2}$, $v_{3}$ and $v_{4}$ of hadrons by the experimental
data of the STAR collaboration and find a good agreement between our theoretical
calculation and the available experiment data at three collision energies.
For the obtained $v_{2}$, $v_{3}$ and $v_{4}$ of constituent quarks
in Au+Au collisions at three collision energies, we discuss some possibly scaling
properties existed in those flows of quarks. 

The paper is organized as follows. In Sec.~\ref{sec:EVC_model},
we derive the analytical formulas of the anisotropic flows of hadrons
in the mechanism of equal-velocity quark combination at hadronization.
In Sec.~\ref{sec:v2}, we test the obtained formulas for $v_{2}$
of hadrons using the experimental data of $\phi$, $\Lambda$, $\Xi^{-}$,
$\Omega^{-}$, $\bar{\Lambda}$, $\bar{\Xi}^{+}$, $\bar{\Omega}^{+}$,
$p$ and $\bar{p}$ in different centralities at $\sqrt{s_{NN}}=$
19.6, 54.4 and 200 GeV. In Sec~\ref{sec:v3}, we show the results
of the $v_{3}$ of identified hadrons $\phi$, $\Lambda$, $\Xi^{-}$,
$\Omega^{-}$, $\bar{\Lambda}$, $\bar{\Xi}^{+}$, $\bar{\Omega}^{+}$,
$p$ and $\bar{p}$ in different centralities at $\sqrt{s_{NN}}=$
19.6, 54.4 and 200 GeV, and make comparison with the available experimental
data. In Sec~\ref{sec:v4}, we show the results of the $v_{4}$ of
identified hadrons. In Sec~\ref{sec:quark properties}, we discuss
the properties of the extracted $v_{2}$, $v_{3}$ and $v_{4}$ of
the up/down and strange quarks. Finally, the summary and discussion
are given in Sec.~\ref{sec:Summary}.

\section{ANISOTROPY FLOW IN QCM WITH EVC\label{sec:EVC_model}}

In this section, we apply a quark combination model (QCM) with EVC approximation
to study the production of hadrons and derive the anisotropic flows
of hadrons at mid-rapidity in heavy-ion collisions. The model
was firstly proposed in Ref. \citep{Song:2017gcz} based on the 
quark number scaling property for the experimental data of $p_T$ spectra of strange hadrons
in $p$-Pb collisions at LHC energy, and can well describe $p_{T}$
spectra and elliptic flow of identified hadrons in $pp$, $p$-Pb
and heavy-ion collisions at RHIC and LHC with collision energies range
from 11.5 to 2760 GeV \citep{Song:2018tpv,Li:2017zuj,Zhang:2018vyr,Song:2019sez,Song:2020kak,Li:2021nhq,Wang:2019fcg}.
We firstly define the inclusive distribution function $f_{h}\left(p_{T},\varphi\right)\equiv dN_{h}/dp_{T}d\varphi$
of particles at mid-rapidity. In the quark combination model with EVC
mechanism, momentum distribution of hadrons is given as the product of those of constituent quarks just before hadronization
\begin{align}
f_{M_{j}}\left(p_{T},\varphi\right) & =\kappa_{M_{j}}f_{q_{1}}\left(x_{1}p_{T},\varphi\right)f_{\bar{q}_{2}}\left(x_{2}p_{T},\varphi\right),\label{eq:fmi_indep}\\
f_{B_{j}}\left(p_{T},\varphi\right) & =\kappa_{B_{j}}f_{q_{1}}\left(x_{1}p_{T},\varphi\right)f_{q_{2}}\left(x_{2}p_{T},\varphi\right)f_{q_{3}}\left(x_{3}p_{T},\varphi\right),\label{eq:fbi_indep}
\end{align}
Here, coefficients $\kappa_{M_{j}}$ and $\kappa_{B_{j}}$ are independent
of the transverse momentum $p_{T}$ and the azimuthal angle $\varphi$.
The $f_{q_{i}}\left(x_{i}p_{T},\varphi\right)$ is the momentum distribution
of quark with flavor $q_{i}$. Momentum fractions $x_{1,2}=m_{1,2}/\left(m_{1}+m_{2}\right)$
for meson $M_{j}\left(q_{1}\bar{q}_{2}\right)$ satisfy the momentum
conservation $x_{1}+x_{2}=1,$ and $x_{1,2,3}=m_{1,2,3}/\left(m_{1}+m_{2}+m_{3}\right)$
for baryon $B_{j}\left(q_{1}q_{2}q_{3}\right)$ satisfies $x_{1}+x_{2}+x_{3}=1$.
The constituent masses of quarks are taken as $m_{u}=m_{d}=0.3$ GeV and $m_{s} = 0.5$ GeV.

The quark distribution function can be expanded as a Fourier series
\begin{equation}
f_{q}\left(p_{T},\varphi\right)=f_{q}\left(p_{T}\right)\biggl[1+2\sum_{k=1}^{\infty}v_{k,q}\left(p_{T}\right)\cos(k\varphi)\biggr].\label{eq:fq}
\end{equation}
We write $f_{q}\left(p_{T}\right)\equiv dN_{q}/dp_{T}$ to denote the part of $\varphi$-independent distribution. The $v_{k}$ is
the Fourier coefficient which is calculated by 
\begin{equation}
v_{k,q}\left(p_{T}\right)=\frac{\int f_{q}\left(p_{T},\varphi\right)\cos(k\varphi)d\varphi}{\int f_{q}\left(p_{T},\varphi\right)d\varphi}.\label{eq:vn}
\end{equation}

Using Eqs. (\ref{eq:fmi_indep}$-$\ref{eq:fq}), we can obtain the
$v_{k}\left(p_{T}\right)$ for the meson $M_{j}\left(q_{1}\bar{q}_{2}\right)$
and baryon $B_{j}\left(q_{1}q_{2}q_{3}\right)$,
\begin{align}
    v_{k,M_{j}} & =\frac{\int f_{M_{j}}\left(p_{T},\varphi\right)\cos(k\varphi) d\varphi}{\int f_{M_{j}}\left(p_{T},\varphi\right)d\varphi}\nonumber \\
 & =\frac{1}{\mathcal{N}_{M_{j}}}\left[v_{k,q_{1}}+v_{k,\overline{q}_{2}}+\sum_{n=1}^{\infty}\sum_{m=1}^{\infty}v_{n,q_{1}}v_{m,\overline{q}_{2}}\left(\delta_{k,m+n}+\delta_{k,n-m}+\delta_{k,m-n}\right)\right],
\end{align}
with 
\begin{equation}
\mathcal{N}_{M_{j}}=1+2\sum_{n=1}^{\infty}v_{n,q_{1}}v_{n,\overline{q}_{2}},
\end{equation}
 and 
\begin{align}
v_{k,B_{j}} & =\frac{\int f_{B_{j}}\left(p_{T},\varphi\right)cosk\varphi d\varphi}{\int f_{B_{j}}\left(p_{T},\varphi\right)d\varphi}\nonumber \\
 & =\frac{1}{\mathcal{N}_{B_{j}}}\left\{ v_{k,q_{1}}+v_{k,q_{2}}+v_{k,q_{3}}+\sum_{n=1}^{\infty}\sum_{m=1}^{\infty}\left(v_{n,q_{1}}v_{m,q_{2}}+v_{n,q_{2}}v_{m,q_{3}}+v_{n,q_{3}}v_{m,q_{1}}\right)\left(\delta_{k,m+n}+\delta_{k,n-m}+\delta_{k,m-n}\right)\right.\nonumber \\
 & +\left.\sum_{n=1}^{\infty}\sum_{m=1}^{\infty}\sum_{l=1}^{\infty}v_{n,q_{1}}v_{m,q_{2}}v_{l,q_{3}}\left(\delta_{k,l+m+n}+\delta_{k,m+n-l}+\delta_{k,l-m+n}+\delta_{k,l+m-n}+\delta_{k,n-l-m}+\delta_{k,m-l-n}+\delta_{k,l-m-n}\right)\right\} ,
\end{align}
with 
\begin{align}
\mathcal{N}_{B_{j}} & =1+2\sum_{m=1}^{\infty}\left(v_{m,q_{2}}v_{m,q_{1}}+v_{m,q_{2}}v_{m,q_{3}}+v_{m,q_{1}}v_{m,q_{3}}\right)\nonumber \\
 & +2\sum_{m=1}^{\infty}\sum_{n=1}^{\infty}\sum_{l=1}^{\infty}v_{m,q_{2}}v_{n,q_{1}}v_{l,q_{3}}\left(\delta_{m,n+l}+\delta_{n,m+l}+\delta_{l,m+n}\right).
\end{align}
In the above formulas, we use the abbreviation $v_{k,q_{i}}$ for $v_{k,q_{i}}\left(x_{i}p_{T}\right)$
with $i=1,2,3$ and $v_{k,\bar{q}_{2}}$ for $v_{k,\bar{q}_{2}}\left(x_{2}p_{T}\right)$.

Taking the power series expansion for $1/\mathcal{N}_{M_{j}}$ and
$1/\mathcal{N}_{B_{j}}$, we obtain 
\begin{align}
v_{k,M_{j}} & =v_{k,q_{1}}\left[1+\sum_{n=1}^{\infty}\frac{v_{n,q_{1}}}{v_{k,q_{1}}}v_{k+n,\overline{q}_{2}}-2\sum_{n=1}^{\infty}v_{n,q_{1}}v_{n,\overline{q}_{2}}\right]\nonumber \\
 & +v_{k,\overline{q}_{2}}\left[1+\sum_{n=1}^{\infty}\frac{v_{n,\overline{q}_{2}}}{v_{k,\overline{q}_{2}}}v_{k+n,q_{1}}-2\sum_{n=1}^{\infty}v_{n,q_{1}}v_{n,\overline{q}_{2}}\right]\nonumber \\
 & +\sum_{n=1}^{k-1}v_{n,q_{1}}v_{k-n,\overline{q}_{2}}+\mathcal{O}(v^{4}),
\end{align}
\begin{align}
 & v_{k,B_{j}}\nonumber \\
 & =v_{k,q_{1}}\left\{ 1+\sum_{n=1}^{\infty}\frac{v_{n,q_{1}}}{v_{k,q_{1}}}\left(v_{k+n,q_{2}}+v_{k+n,q_{3}}\right)-2\sum_{n=1}^{\infty}\left(v_{n,q_{2}}v_{n,q_{1}}+v_{n,q_{2}}v_{n,q_{3}}+v_{n,q_{1}}v_{n,q_{3}}\right)\right.\nonumber \\
 & \left.+\sum_{n=1}^{\infty}\sum_{m=1}^{\infty}\frac{v_{n,q_{1}}}{v_{k,q_{1}}}v_{m,q_{2}}\left(v_{m+n+k,q_{3}}+v_{m+n-k,q_{3}}\right)\right\} \nonumber \\
 & +v_{k,q_{2}}\left\{ 1+\sum_{n=1}^{\infty}\frac{v_{n,q_{2}}}{v_{k,q_{2}}}\left(v_{k+n,q_{3}}+v_{k+n,q_{1}}\right)-2\sum_{n=1}^{\infty}\left(v_{n,q_{2}}v_{n,q_{1}}+v_{n,q_{2}}v_{n,q_{3}}+v_{n,q_{1}}v_{n,q_{3}}\right)\right.\nonumber \\
 & \left.+\sum_{n=1}^{\infty}\sum_{m=1}^{\infty}\frac{v_{n,q_{2}}}{v_{k,q_{2}}}v_{m,q_{3}}\left(v_{m+n+k,q_{1}}+v_{m+n-k,q_{1}}\right)\right\} \nonumber \\
 & +v_{k,q_{3}}\left\{ 1+\sum_{n=1}^{\infty}\frac{v_{n,q_{3}}}{v_{k,q_{3}}}\left(v_{k+n,q_{1}}+v_{k+n,q_{2}}\right)-2\sum_{n=1}^{\infty}\left(v_{n,q_{2}}v_{n,q_{1}}+v_{n,q_{2}}v_{n,q_{3}}+v_{n,q_{1}}v_{n,q_{3}}\right)\right.\nonumber \\
 & \left.+\sum_{n=1}^{\infty}\sum_{m=1}^{\infty}\frac{v_{n,q_{3}}}{v_{k,q_{3}}}v_{m,q_{1}}\left(v_{m+n+k,q_{2}}+v_{m+n-k,q_{2}}\right)\right\} \nonumber \\
 & +\sum_{n=1}^{k-1}\left(v_{n,q_{1}}v_{k-n,q_{2}}+v_{n,q_{2}}v_{k-n,q_{3}}+v_{n,q_{3}}v_{k-n,q_{1}}\right)+\sum_{n,m=1}^{m+n\leq k-1}v_{n,q_{1}}v_{m,q_{2}}v_{k-m-n,q_{3}}+\mathcal{O}(v^{4}).
\end{align}
We see that the anisotropic flow $v_{k}$ of hadrons in the quark combination
mechanism comes from the two contribution classes of quark flows.
The first part is the summation of $v_{k}$ of constituent (anti-)quarks
with a modification factor. The modification factor is close to one, 
and the deviation comes from $v_{n}^{2}$ of (anti-)quarks and $v_{n}^{2}v_{m}^{2}$
of quarks (only in the baryon case). The second part is the product
of the low-order flows of two quarks and those of three quarks (only
in baryon case) with a global constraint that the summation of order
index of quarks flows should equal to $k$. 
\begin{widetext}
Because only low-order flows of hadrons are measured and practically
known in relativistic heavy-ion collisions, we should further simplify
the formulas of the anisotropic flows of hadrons by virtue of the available
experimental measurements to make them calculable and predictive.
In this paper, we study the flows of hadrons at mid-rapidity in relativistic
heavy-ion collisions. First, we ignore the influence of quarks $v_{1}$
because it is very small at mid-rapidity by the estimation $v_{1,q}\sim v_{1,h}/n\leq10^{-3}$
from the measurements of $v_{1}$ of hadrons in heavy-ion collisions
at RHIC and LHC energies \citep{STAR:2008jgm,ALICE:2013xri}. Second,
the 2$-$$5^{th}$ flows of quarks estimated from hadrons by NCQ are
about $v_{2/3/4\text{/5,\ensuremath{q}}}\sim10^{-2}$ \citep{PHENIX:2006dpn,STAR:2015gge,ALICE:2014wao,ALICE:2018yph,STAR:2013ayu},
the magnitude of the high power terms such as $\left(v_{n,q}\right)^{4,5}$
is very small and could be safely neglected. Third, the experimental
data of the anisotropic flows of hadrons indicate a scaling relation
$v_{n,h}\left(p_{T}\right)\sim v_{2,h}^{n/2}\left(p_{T}\right)$ \citep{STAR:2004jwm,STAR:2007afq,STAR:2003xyj}.
The scaling relation has been studied theoretically in both the transport
models \citep{Kolb:2004gi,Chen:2004dv,Zhang:2015skc,Konchakovski:2012yg}
and the hydrodynamic model \citep{Gardim:2012yp,Borghini:2005kd,Gombeaud:2009ye}.
We assume this scaling relation also roughly exists at quark level,
i.e., 
\begin{equation}
v_{n,q}=a_{n}v_{2,q}^{n/2}.\label{eq:vnq_scaling}
\end{equation}

Then flows of hadrons in QCM can be greatly simplified. 
\begin{align}
v_{k,M_{j}} & \approx v_{k,q_{1}}\left[1+\sum_{n=2}^{\infty}\left(\frac{a_{k+n}}{a_{k}a_{n}}-2\right)v_{n,q_{1}}v_{n,\overline{q}_{2}}\right]+v_{k,\overline{q}_{2}}\left[1+\sum_{n=2}^{\infty}\left(\frac{a_{k+n}}{a_{k}a_{n}}-2\right)v_{n,q_{1}}v_{n,\overline{q}_{2}}\right]\nonumber \\
 & +\sum_{n=2}^{k-2}v_{n,q_{1}}v_{k-n,\overline{q}_{2}},
\end{align}
\begin{align}
 & v_{k,B_{j}}\nonumber \\
 & \approx v_{k,q_{1}}\left\{ 1+\sum_{n=2}^{\infty}\left(\frac{a_{k+n}}{a_{k}a_{n}}-2\right)\left(v_{n,q_{1}}v_{n,q_{2}}+v_{n,q_{1}}v_{n,q_{3}}\right)-2\sum_{n=2}^{\infty}v_{n,q_{2}}v_{n,q_{3}}+\sum_{n,m=2}^{m+n\geq k+2}\frac{v_{n,q_{1}}v_{m,q_{2}}}{v_{k,q_{1}}}v_{m+n-k,q_{3}}\right\} \nonumber \\
 & +v_{k,q_{2}}\left\{ 1+\sum_{n=2}^{\infty}\left(\frac{a_{k+n}}{a_{k}a_{n}}-2\right)\left(v_{n,q_{1}}v_{n,q_{2}}+v_{n,q_{2}}v_{n,q_{3}}\right)-2\sum_{n=2}^{\infty}v_{n,q_{1}}v_{n,q_{3}}+\sum_{n,m=2}^{m+n\geq k+2}\frac{v_{n,q_{2}}v_{m,q_{3}}}{v_{k,q_{2}}}v_{m+n-k,q_{1}}\right\} \nonumber \\
 & +v_{k,q_{3}}\left\{ 1+\sum_{n=2}^{\infty}\left(\frac{a_{k+n}}{a_{k}a_{n}}-2\right)\left(v_{n,q_{1}}v_{n,q_{3}}+v_{n,q_{2}}v_{n,q_{3}}\right)-2\sum_{n=2}^{\infty}v_{n,q_{2}}v_{n,q_{1}}+\sum_{n,m=2}^{m+n\geq k+2}\frac{v_{n,q_{1}}v_{m,q_{3}}}{v_{k,q_{3}}}v_{m+n-k,q_{2}}\right\} \nonumber \\
 & +\sum_{n=2}^{k-2}\left(v_{n,q_{1}}v_{k-n,q_{2}}+v_{n,q_{2}}v_{k-n,q_{3}}+v_{n,q_{3}}v_{k-n,q_{1}}\right)+\sum_{n,m=2}^{m+n\leq k-2}v_{n,q_{1}}v_{m,q_{2}}v_{k-m-n,q_{3}}.
\end{align}

Furthermore, based on the scaling assumption Eq. ($\ref{eq:vnq_scaling}$)
and the measured $v_{k}$ of hadrons in relativistic heavy-ion collisions
\citep{PHENIX:2006dpn,STAR:2015gge,STAR:2022ncy,ALICE:2014wao,ALICE:2018yph},
we have estimate $v_{n+1,q_{1}}v_{n+1,q_{2}}/v_{n,q_{1}}v_{n,q_{2}}\sim10^{-2}$.
Therefore, we can neglect the correction terms with index $n,m>2$
in the modification factor after $v_{k,q_{i}}$ and finally obtain 
\begin{align}
v_{k,M_{j}} & \approx v_{k,q_{1}}\left[1+\left(\frac{a_{k+2}}{a_{k}a_{2}}-2\right)v_{2,q_{1}}v_{2,\overline{q}_{2}}\right]+v_{k,\overline{q}_{2}}\left[1+\left(\frac{a_{k+2}}{a_{k}a_{2}}-2\right)v_{2,q_{1}}v_{2,\overline{q}_{2}}\right]\nonumber \\
 & +\sum_{n=2}^{k-2}v_{n,q_{1}}v_{k-n,\overline{q}_{2}},\label{eq:vkm_final}
\end{align}
\begin{align}
 & v_{k,B_{j}}\nonumber \\
 & \approx v_{k,q_{1}}\left\{ 1+\left(\frac{a_{k+2}}{a_{k}a_{2}}-2\right)\left(v_{2,q_{1}}v_{2,q_{2}}+v_{2,q_{1}}v_{2,q_{3}}\right)-2v_{2,q_{2}}v_{2,q_{3}}+\sum_{n,m=2}^{m+n\geq k+2}\frac{v_{n,q_{1}}v_{m,q_{2}}}{v_{k,q_{1}}}v_{m+n-k,q_{3}}\right\} \nonumber \\
 & +v_{k,q_{2}}\left\{ 1+\left(\frac{a_{k+2}}{a_{k}a_{2}}-2\right)\left(v_{2,q_{1}}v_{2,q_{2}}+v_{2,q_{2}}v_{2,q_{3}}\right)-2v_{2,q_{1}}v_{2,q_{3}}+\sum_{n,m=2}^{m+n\geq k+2}\frac{v_{n,q_{2}}v_{m,q_{3}}}{v_{k,q_{2}}}v_{m+n-k,q_{1}}\right\} \nonumber \\
 & +v_{k,q_{3}}\left\{ 1+\left(\frac{a_{k+2}}{a_{k}a_{2}}-2\right)\left(v_{2,q_{1}}v_{2,q_{3}}+v_{2,q_{2}}v_{2,q_{3}}\right)-2v_{2,q_{2}}v_{2,q_{1}}+\sum_{n,m=2}^{m+n\geq k+2}\frac{v_{n,q_{1}}v_{m,q_{3}}}{v_{k,q_{3}}}v_{m+n-k,q_{2}}\right\} \nonumber \\
 & +\sum_{n=2}^{k-2}\left(v_{n,q_{1}}v_{k-n,q_{2}}+v_{n,q_{2}}v_{k-n,q_{3}}+v_{n,q_{3}}v_{k-n,q_{1}}\right)+\sum_{n,m=2}^{m+n\leq k-2}v_{n,q_{1}}v_{m,q_{2}}v_{k-m-n,q_{3}}.\label{eq:vkb_final}
\end{align}
\end{widetext}

\section{THE ELLIPTIC FLOW OF IDENTIFIED HADRONS\label{sec:v2} }

In this section, we use the above EVC model to study the $p_{T}$ dependence
of the $v_{2}$ for the identified hadrons by systematically analyzing
the experimental data of $\phi$, $\Lambda$, $\Xi^{-}$, $\Omega^{-}$,
$\bar{\Lambda}$, $\bar{\Xi}^{+}$, $\bar{\Omega}^{+}$, $p$ and
$\bar{p}$ at mid-rapidity ($\left|y\right|$<1.0) in different centralities
(0-10\%, 0-80\%, 0-30\%, 10-40\%, 40-80\% and 30-80\%) in Au+Au collisions
at $\sqrt{s_{NN}}=$ 19.6, 54.4 and 200 GeV \citep{Dixit:2022geb,STAR:2015rxv,STAR:2022tfp,Sharma:2023dkg,STAR:2022ncy,STAR:2015gge,STAR:2013ayu}.  According to Eqs. (\ref{eq:vkm_final}) and (\ref{eq:vkb_final}),
$v_{2}$ of hadrons are written as 
\begin{align}
v_{2,M_{j}} & \approx v_{2,q_{1}}\left[1+\left(a_{4}-2\right)v_{2,q_{1}}v_{2,\bar{q}_{2}}\right]+v_{2,\bar{q}_{2}}\left[1+\left(a_{4}-2\right)v_{2,q_{1}}v_{2,\bar{q}_{2}}\right],\label{eq:v2_Mi}\\
v_{2,B_{j}} & \approx v_{2,q_{1}}\left[1+\left(a_{4}-2\right)\left(v_{2,q_{1}}v_{2,q_{2}}+v_{2,q_{1}}v_{2,q_{3}}\right)-v_{2,q_{2}}v_{2,q_{3}}\right]\nonumber \\
 & +v_{2,q_{2}}\left[1+\left(a_{4}-2\right)\left(v_{2,q_{1}}v_{2,q_{2}}+v_{2,q_{2}}v_{2,q_{3}}\right)-v_{2,q_{1}}v_{2,q_{3}}\right]\nonumber \\
 & +v_{2,q_{3}}\left[1+\left(a_{4}-2\right)\left(v_{2,q_{1}}v_{2,q_{3}}+v_{2,q_{2}}v_{2,q_{3}}\right)-v_{2,q_{2}}v_{2,q_{1}}\right].\label{eq:v2_Bi}
\end{align}

We observe that the $v_{2}$ of the hadrons is simply the summation
of $v_{2}$ of constituent quarks and/or that of anti-quarks multiplied
by a modification factor close to one. Coefficient $a_{4}$ in the modification
term is approximately 2 according to the calculation of the multiphase
transport (AMPT) model and our preliminary analysis of experimental
data \citep{Zhang:2015skc,STAR:2007afq}. The magnitude of $v_{2,q}^{2}$
is about $10^{-3}$ according to NCQ estimations from experimental
data of $v_{2}$ of hadrons \citep{PHENIX:2006dpn,STAR:2015gge,ALICE:2014wao,ALICE:2018yph,STAR:2013ayu}.
In the studied collision energies and centralities, the magnitude
of $(a_{4}-2)v_{2,q}^{2}$ in the modification term is approximately
less than 0.7\% in the central and semi-central collisions, and less
than about 2\% in the peripheral collisions. Therefore, we neglect
the effect of these small modifications and express $v_{2}$ of hadrons
as the simplest form 
\begin{align}
v_{2,M_{j}}\left(p_{T}\right) & \approx v_{2,q_{1}}\left(x_{1}p_{T}\right)+v_{2,\bar{q}_{2}}\left(x_{2}p_{T}\right),\label{eq:v2,Mi_simp}\\
v_{2,B_{j}}\left(p_{T}\right) & \approx v_{2,q_{1}}\left(x_{1}p_{T}\right)+v_{2,q_{2}}\left(x_{2}p_{T}\right)+v_{2,q_{3}}\left(x_{3}p_{T}\right).\label{eq:v2,Bi_simp}
\end{align}

Using Eqs. (\ref{eq:v2,Mi_simp}) and (\ref{eq:v2,Bi_simp}), we obtain
the analytic formulas for the $v_{2}$ of the following hadrons
\begin{align}
v_{2,p}\left(p_{T}\right) & =2v_{2,u}\left(p_{T}/3\right)+v_{2,d}\left(p_{T}/3\right),\label{eq:v2,p}\\
v_{2,\Omega}\left(p_{T}\right) & =3v_{2,s}\left(p_{T}/3\right),\label{eq:v2,omg}\\
v_{2,\phi}\left(p_{T}\right) & =v_{2,s}\left(p_{T}/2\right)+v_{2,\overline{s}}\left(p_{T}/2\right),\label{eq:v2,phi}\\
v_{2,\Lambda}\left(p_{T}\right) & =v_{2,u}\bigl(\frac{1}{2+r}p_{T}\bigr)+v_{2d}\bigl(\frac{1}{2+r}p_{T}\bigr)+v_{2,s}\bigl(\frac{r}{2+r}p_{T}\bigr),\label{eq:v2,lamb}\\
v_{2,\Xi}\left(p_{T}\right) & =v_{2,d}\bigl(\frac{1}{1+2r}p_{T}\bigr)+2v_{2,s}\bigl(\frac{r}{1+2r}p_{T}\bigr)\text{,}\label{eq:v2,xi}
\end{align}
where the factor $r=m_{s}/m_{u}=1.667$. Here, we see the simple summation
rule in $v_{2}$ of hadrons which is convenient for the experimental
test. Considering the collision energies ($\sqrt{s_{NN}}\geq19.6$
GeV) and kinetic region ($\left|y\right|<0.5$), we take the approximation
of isospin symmetry between up and down quarks, $v_{2,u}=v_{2,d}$
and $v_{2,\bar{u}}=v_{2,\bar{d}}$, and assume the strangeness neutrality,
i.e., $v_{2,s}=v_{2,\text{\ensuremath{\bar{s}}}}$. Formulas of proton,
$\phi$, $\Lambda$ and $\Xi$ are further simplified
\begin{align}
v_{2,p}\left(p_{T}\right) & =3v_{2,u}\left(p_{T}/3\right),\label{eq:v2,p-1}\\
v_{2,\phi}\left(p_{T}\right) & =2v_{2,s}\left(p_{T}/2\right),\label{eq:v2,phi-1}\\
v_{2,\Lambda}\left(p_{T}\right) & =2v_{2,u}\bigl(\frac{1}{2+r}p_{T}\bigr)+v_{2,s}\bigl(\frac{r}{2+r}p_{T}\bigr),\label{eq:v2,lamb-1}\\
v_{2,\Xi}\left(p_{T}\right) & =v_{2,u}\bigl(\frac{1}{1+2r}p_{T}\bigr)+2v_{2,s}\bigl(\frac{r}{1+2r}p_{T}\bigr)\text{.}\label{eq:v2,xi-1}
\end{align}

From Eqs. (\ref{eq:v2,omg}) and (\ref{eq:v2,p-1})-(\ref{eq:v2,xi-1}),
we see that $v_{2}$ of $\phi$, $p$, $\Lambda$, $\Xi$ and $\Omega$
are correlated at the quark level by $v_{2,u}$ and $v_{2,s}$. Now,
we use the experimental data of $v_{2}$ of hadrons measured by the STAR
collaboration at RHIC \citep{Dixit:2022geb,STAR:2015rxv,STAR:2022tfp,Sharma:2023dkg,STAR:2022ncy,STAR:2015gge,STAR:2013ayu}
to test these relationships. First, we use the experimental data of
the $v_{2}$ of $\phi$ to extract $v_{2,s}$ by Eq. (\ref{eq:v2,phi-1}).
Because the $p_{T}$ coverage of the experimental data for the $v_{2}$ of
$p$ and $\bar{p}$ is small, we use the experimental data of the
$v_{2}$ of $\Lambda$ and $\bar{\Lambda}$ to extract the $v_{2,u}$
and $v_{2,\bar{u}}$ by Eq. (\ref{eq:v2,lamb-1}), respectively. $v_{2}$
of (anti-)quarks in the extraction process is parameterized as 
\begin{equation}
v_{2,q}\left(p_{T}\right)=a_{q}\exp\left[-\frac{p_{T}}{b_{q}}-c_{q}\exp\left(-\frac{p_{T}}{d_{q}}\right)\right],\label{eq:v2q}
\end{equation}
where $a_{q},$ $b_{q},$ $c_{q}$ and $d_{q}$ are parameters that
control the shape of the $v_{2}$ of quarks. Second, we use the extracted
$v_{2}$ of quarks and those of anti-quarks to calculate $v_{2}$
of $p$, $\Xi$, $\Omega$ and their anti-particles to test the formulas
of our model in Eqs. (\ref{eq:v2,omg}) and (\ref{eq:v2,p-1})-(\ref{eq:v2,xi-1}).

\begin{figure}
\includegraphics[width=0.6\linewidth,viewport=0bp 00bp 520bp 480bp]{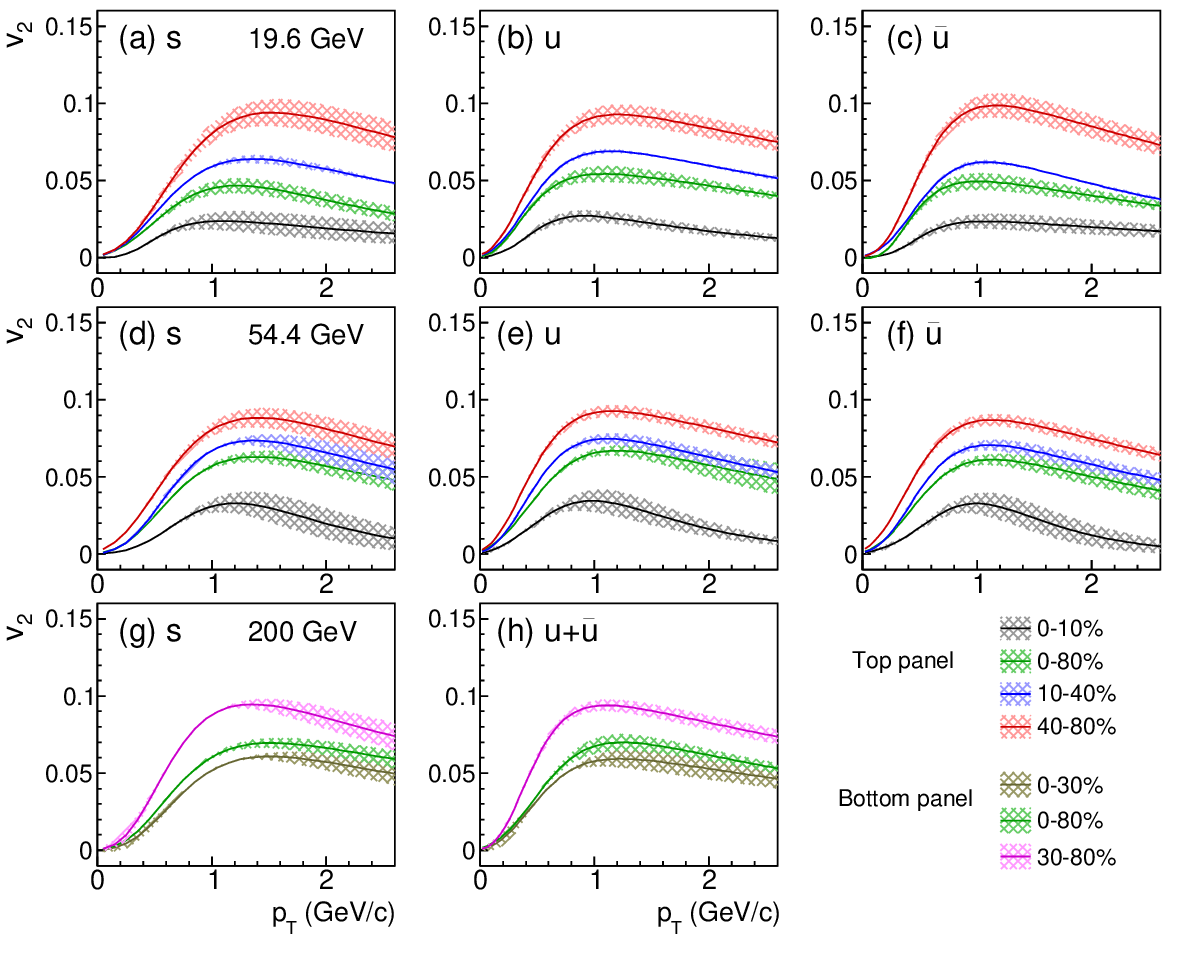}\caption{$v_{2}$ of quarks as a function of $p_{T}$ in different centralities
in Au+Au collisions at $\sqrt{s_{NN}}=$19.6, 54.4 and 200 GeV. \label{fig:fig1}}
\end{figure}

In Fig.~\ref{fig:fig1}, $v_{2}$ of up and strange (anti-)quarks
as the function of $p_{T}$ at mid-rapidity ($\left|y\right|$<1.0)
in different centralities in Au+Au collisions at $\sqrt{s_{NN}}=$19.6,
54.4 and 200 GeV are shown. The shadow areas show the statistical
uncertainties, which are passed from the statistical uncertainties
of the experimental data of $\phi$, $\Lambda$ and $\bar{\Lambda}$.

\begin{figure}
\includegraphics[width=0.9\linewidth,viewport=0bp 0bp 556bp 242bp]{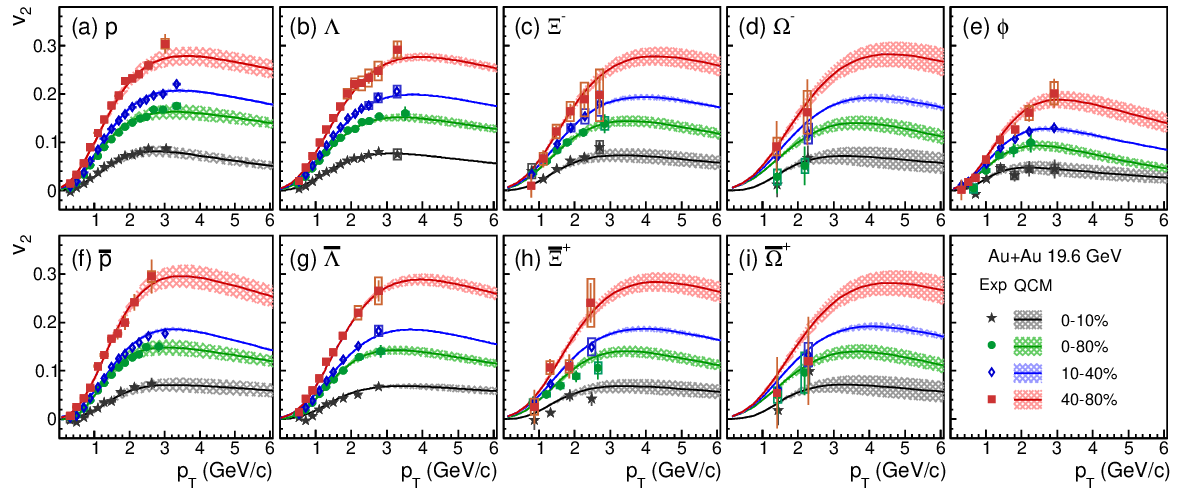}\caption{$v_{2}$ as a function of $p_{T}$ at mid-rapidity($\left|y\right|<1.0$)
for 0-10\%, 0-80\%, 10-40\% and 40-80\% centralities in Au+Au collisions
at $\sqrt{s_{NN}}=$19.6 GeV. Symbols are the experimental data \citep{Dixit:2022geb,STAR:2015rxv,STAR:2013ayu}
and lines are model results. \label{fig:fig2}}
\end{figure}

\begin{figure}
\includegraphics[width=0.9\linewidth,viewport=0bp 0bp 556bp 242bp]{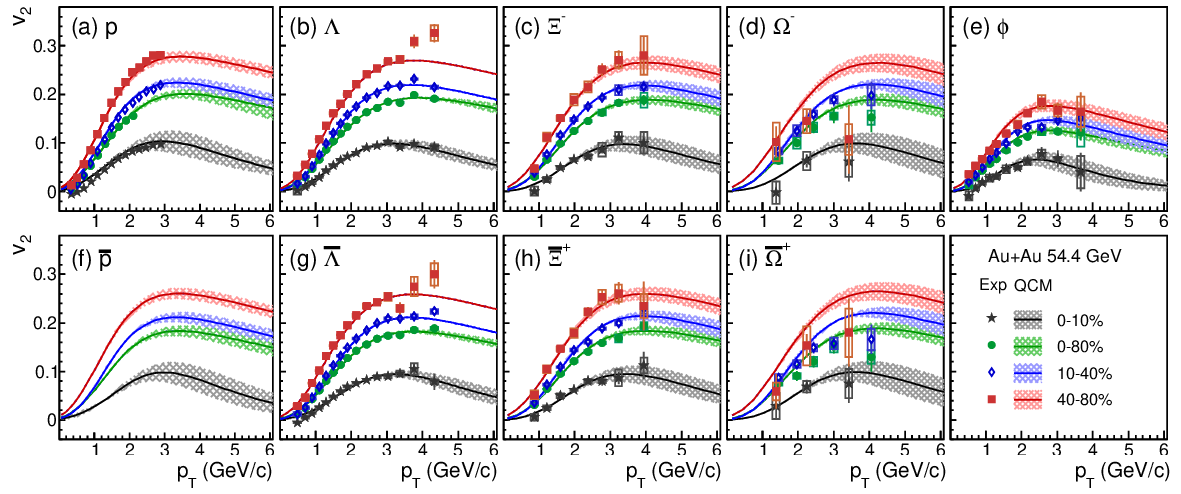}\caption{$v_{2}$ as a function of $p_{T}$ at mid-rapidity($\left|y\right|<1.0$)
for 0-10\%, 0-80\%, 10-40\% and 40-80\% centralities in Au+Au collisions
at $\sqrt{s_{NN}}=$54.4 GeV. Symbols are the experimental data \citep{STAR:2022tfp,Sharma:2023dkg}
and lines are model results.\label{fig:fig3}}
\end{figure}

\begin{figure}
\includegraphics[width=0.65\linewidth,viewport=0bp 00bp 550bp 350bp]{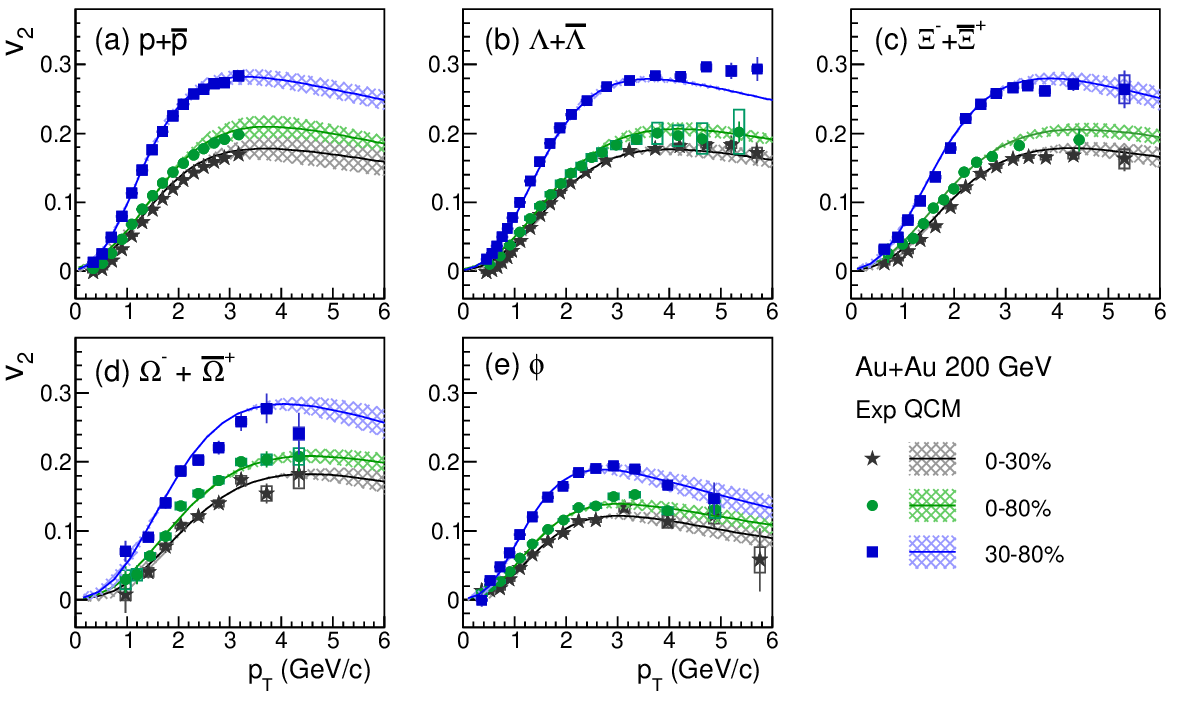}\caption{$v_{2}$ as a function of $p_{T}$ at mid-rapidity($\left|y\right|<1.0$)
for 0-30\%, 0-80\% and 30-80\% centralities in Au+Au collisions at
$\sqrt{s_{NN}}=$200 GeV. Symbols are the experimental data \citep{STAR:2022ncy,STAR:2015gge}
and lines are model results.\label{fig:fig4}}
\end{figure}

In Figs.~\ref{fig:fig2}$-$\ref{fig:fig4}, we show $v_{2}$ of the
identified hadrons $\phi$, $\Lambda$, $\Xi^{-}$, $\Omega^{-}$,
$\bar{\Lambda}$, $\bar{\Xi}^{+}$, $\bar{\Omega}^{+}$, $p$ and
$\bar{p}$ at mid-rapidity and compare them with the experimental data in Au + Au collisions at $\sqrt{s_{NN}}=$19.6, 54.4 and 200 GeV in different centralities (0-10\%, 0-80\%, 0-30\%, 10-40\%, 40-80\% and 30-80\%)
\citep{Dixit:2022geb,STAR:2015rxv,STAR:2022tfp,Sharma:2023dkg,STAR:2022ncy,STAR:2015gge}.
Theoretical results are shown as lines and data are shown as symbols.
Note that the experimental data of $\Lambda$, $\bar{\Lambda}$ and
$\phi$ are used to extract $v_{2}$ of quarks while results of other
identified hadrons $\Xi^{-}$, $\bar{\Xi}^{+}$, $\Omega^{-}$, $\bar{\Omega}^{+}$,
$p$ and $\bar{p}$ are theoretical predictions. By comparing
between theoretical results and experimental data, we see that our
predictions of $p$, $\Xi$, $\Omega$ are generally in good agreement
with the experimental data in the available $p_{T}$ range.

\section{THE TRIANGULAR FLOW OF IDENTIFIED HADRONS\label{sec:v3}}

In this section, we study the $v_{3}$ of identified hadrons $\phi$,
$\Lambda$, $\Xi^{-}$, $\Omega^{-}$, $\bar{\Lambda}$, $\bar{\Xi}^{+}$,
$\bar{\Omega}^{+}$, $p$ and $\bar{p}$ at mid-rapidity ($\left|y\right|$<1.0)
in different centralities in Au+Au collisions at $\sqrt{s_{NN}}=$19.6,
54.4, and 200 GeV \citep{Dixit:2022geb,STAR:2022ncy,STAR:2022tfp}.
The $v_{3}$ of hadrons is related to the initial geometry fluctuations
\citep{Alver:2010gr}. In the quark combination mechanism, $v_{3}$ of
hadrons can be attributed to those of (anti-)quarks just before hadronization.
According to Eqs. (\ref{eq:vkm_final}) and (\ref{eq:vkb_final}),
$v_{3}$ of hadrons can be written as 
\begin{flushleft}
\begin{align}
v_{3,M_{j}} & \approx v_{3,q_{1}}\left[1+\left(\frac{a_{5}}{a_{3}}-2\right)v_{2,q_{1}}v_{2,\bar{q}_{2}}\right]+v_{3,\bar{q}_{2}}\left[1+\left(\frac{a_{5}}{a_{3}}-2\right)v_{2,q_{1}}v_{2,\bar{q}_{2}}\right],\label{eq:v3,Mi}\\
v_{3,B_{j}} & \approx v_{3,q_{1}}\left[1+\left(\frac{a_{5}}{a_{3}}-2\right)\left(v_{2,q_{1}}v_{2,q_{2}}+v_{2,q_{1}}v_{2,q_{3}}\right)\right]+v_{3,q_{2}}\left[1+\left(\frac{a_{5}}{a_{3}}-2\right)\left(v_{2,q_{1}}v_{2,q_{2}}+v_{2,q_{2}}v_{2,q_{3}}\right)\right]\nonumber \\
 & +v_{3,q_{3}}\left[1+\left(\frac{a_{5}}{a_{3}}-2\right)\left(v_{2,q_{1}}v_{2,q_{3}}+v_{2,q_{2}}v_{2,q_{3}}\right)\right].\label{eq:v3,Bi}
\end{align}
\end{flushleft}

We see that $v_{3}$ of hadrons can be expressed as the summation
of $v_{3}$ of quarks and that of anti-quarks multiplied by a modification
factor about one. The leading modification term is also the order
of $v_{2,q}^{2}$, which is quite similar to that in $v_{2}$ of
hadrons. According to the NCQ estimation of the 2-$5^{th}$ flow of
hadrons \citep{PHENIX:2006dpn,STAR:2015gge,ALICE:2014wao,ALICE:2018yph,STAR:2013ayu},
the $a_{5}/a_{3}$ is about 2.5 \citep{Kolb:2004gi,Chen:2004dv,STAR:2004jwm,STAR:2007afq,Bairathi:2016dob}.
Therefore, the magnitude of the $(a_{5}/a_{3}-2)v_{2,q}^{2}$ is about
$10^{-3}$ and its influence can be neglected in the formula, and we finally
obtain 
\begin{align}
v_{3,M_{j}}\left(p_{T}\right) & \approx v_{3,q_{1}}\left(x_{1}p_{T}\right)+v_{3,\bar{q}_{2}}\left(x_{2}p_{T}\right),\label{eq:v3,Mi_simp}\\
v_{3,B_{j}}\left(p_{T}\right) & \approx v_{3,q_{1}}\left(x_{1}p_{T}\right)+v_{3,q_{2}}\left(x_{2}p_{T}\right)+v_{3,q_{3}}\left(x_{3}p_{T}\right).\label{eq:v3,Bi_simp}
\end{align}
Applying Eqs. (\ref{eq:v3,Mi_simp}$-$\ref{eq:v3,Bi_simp}), we
obtain
\begin{align}
v_{3,p}\left(p_{T}\right) & =3v_{3,u}\left(p_{T}/3\right),\label{eq:v3,p}\\
v_{3,\Omega}\left(p_{T}\right) & =3v_{3,s}\left(p_{T}/3\right),\label{eq:v3,omg}\\
v_{3,\phi}\left(p_{T}\right) & =v_{3,s}\left(p_{T}/2\right)+v_{3,\overline{s}}\left(p_{T}/2\right),\label{eq:v3,phi}\\
v_{3,\Lambda}\left(p_{T}\right) & =2v_{3,u}\bigl(\frac{1}{2+r}p_{T}\bigr)+v_{3,s}\bigl(\frac{r}{2+r}p_{T}\bigr),\label{eq:v3,lamb}\\
v_{3,\Xi}\left(p_{T}\right) & =2v_{3,s}\bigl(\frac{r}{1+2r}p_{T}\bigr)+v_{3,u}\bigl(\frac{1}{1+2r}p_{T}\bigr),\label{eq:v3,xi}
\end{align}
here, we also apply the isospin symmetry $v_{3,u}=v_{3,d}$ and charge
conjugation symmetry $v_{3,s}=v_{3,\bar{s}}$ at mid-rapidity in heavy-ion
collisions at high RHIC energies. 

\begin{figure}
\includegraphics[width=0.65\linewidth,viewport=0bp 00bp 520bp 480bp]{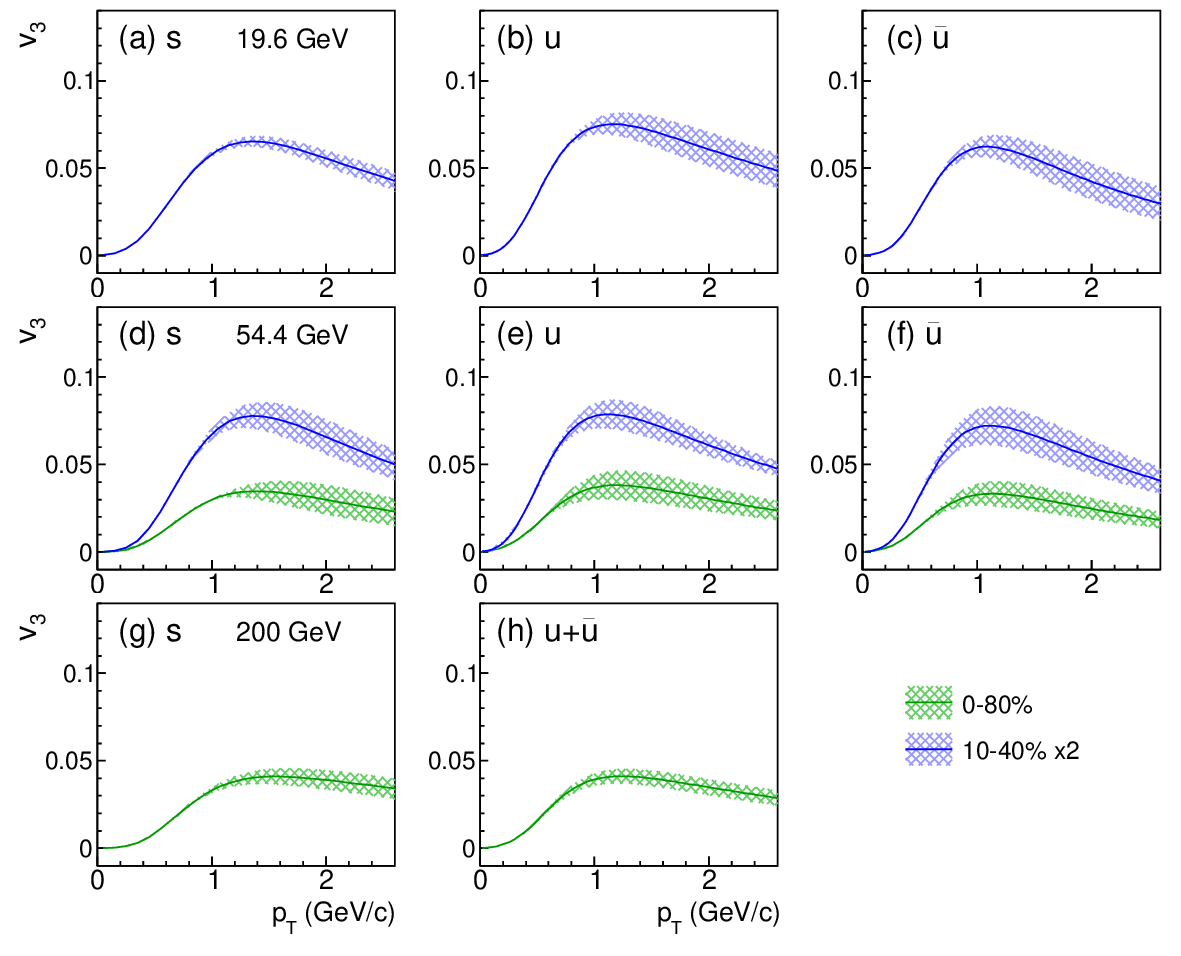}\caption{$v_{3}$ of quarks as a function of $p_{T}$ in different centralities
in Au+Au collisions at $\sqrt{s_{NN}}=$19.6, 54.4 and 200 GeV. \label{fig:fig5}}
\end{figure}

\begin{figure}
\includegraphics[width=0.9\linewidth,viewport=0bp 0bp 556bp 242bp]{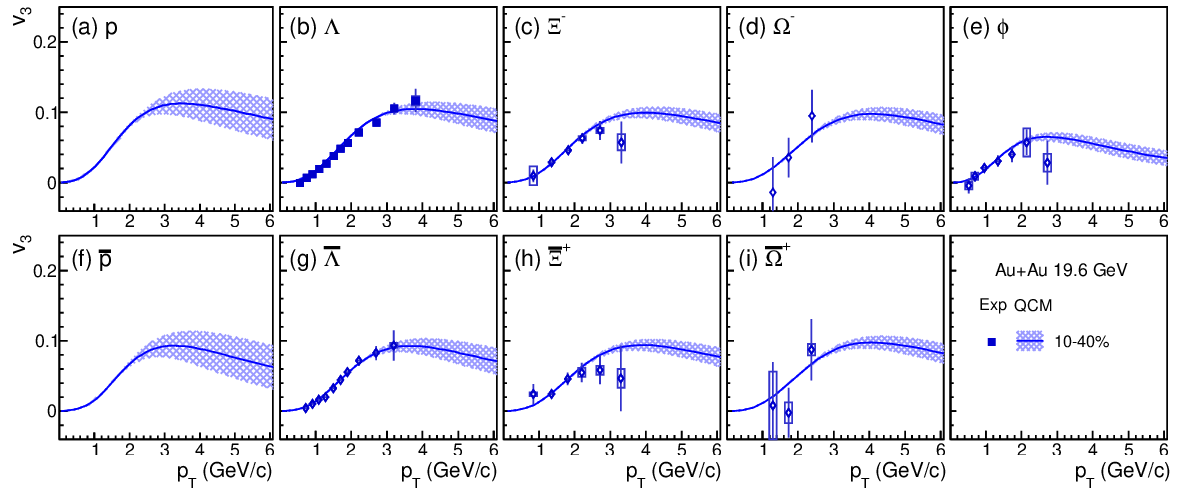}\caption{$v_{3}$ as a function of $p_{T}$ at mid-rapidity($\left|y\right|<1.0$)
for 10-40\% centrality in Au+Au collisions at $\sqrt{s_{NN}}=$19.6
GeV. Symbols are the experimental data \citep{Dixit:2022geb} and
lines are model results. \label{fig:fig6}}
\end{figure}

\begin{figure}
\includegraphics[width=0.9\linewidth,viewport=0bp 0bp 556bp 242bp]{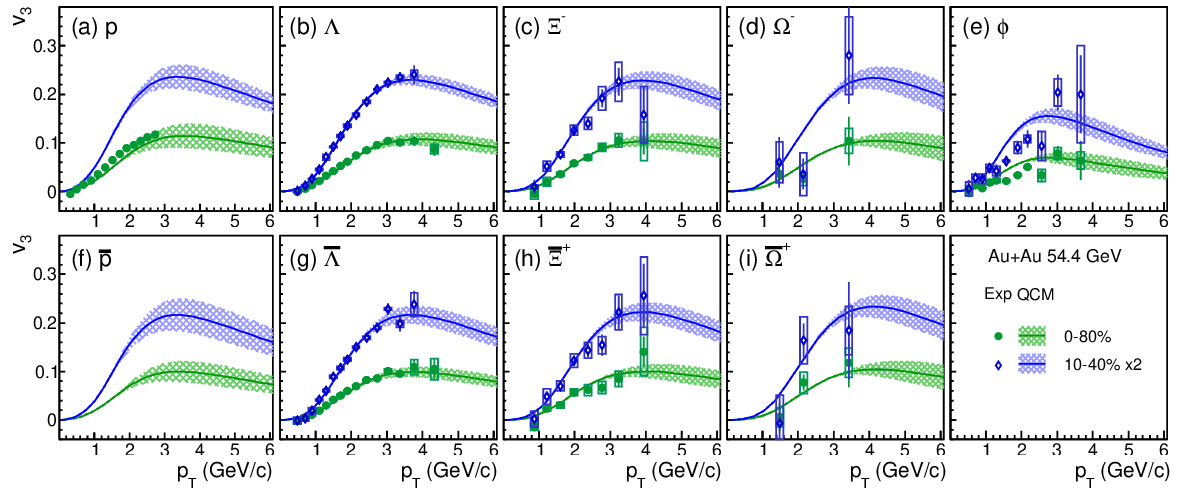}\caption{$v_{3}$ as a function of $p_{T}$ at mid-rapidity($\left|y\right|<1.0$)
for 0-80\% and 10-40\% centralities in Au+Au collisions at $\sqrt{s_{NN}}=$54.4
GeV. Symbols are the experimental data \citep{STAR:2022tfp} and lines
are model results. \label{fig:fig7}}
\end{figure}

\begin{figure}
\includegraphics[width=0.65\linewidth,viewport=0bp 00bp 550bp 350bp]{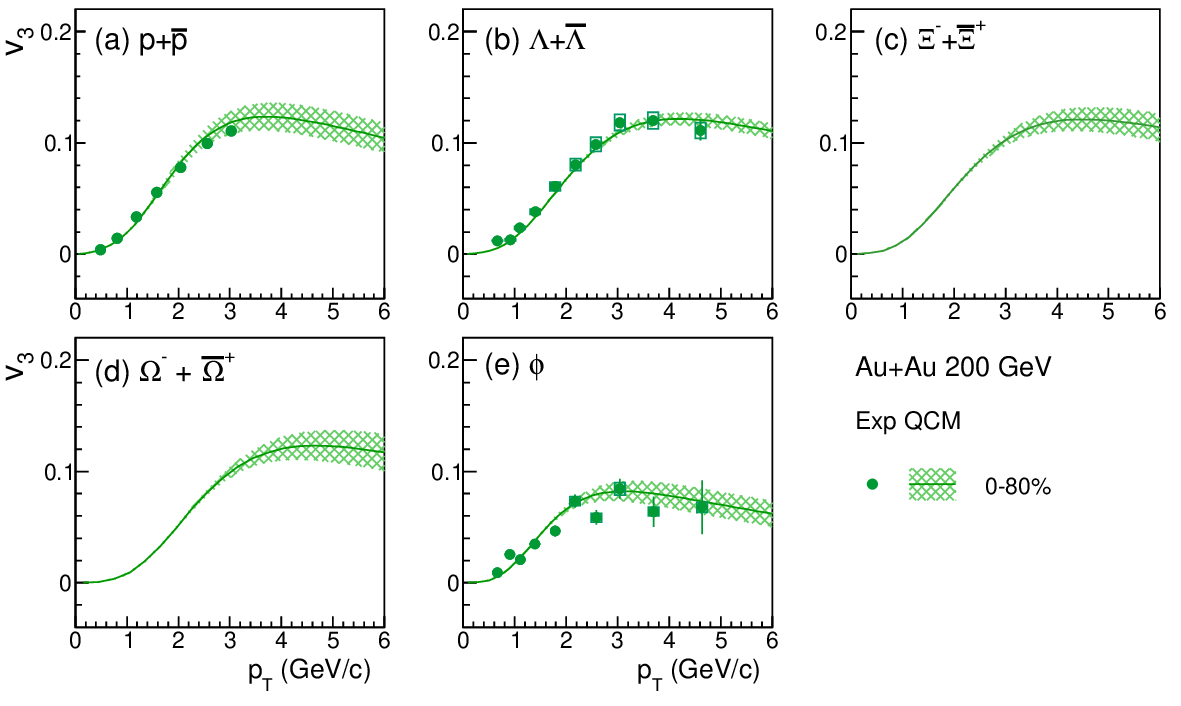}\caption{$v_{3}$ as a function of $p_{T}$ at mid-rapidity($\left|y\right|<1.0$)
for 0-80\% centrality in Au+Au collisions at $\sqrt{s_{NN}}=$200
GeV. Symbols are the experimental data \citep{STAR:2022ncy} and lines
are model results. \label{fig:fig8}}
\end{figure}

We adopt the procedure which is similar to that in $v_{2}$ of hadrons
in previous section to test formulas of $v_{3}$ of hadrons. First,
we take the formula in Eq. (\ref{eq:v2q}) to parameterize the $v_{3}$
of (anti-)quarks and extract $v_{3,s}$, $v_{3,u}$ and $v_{3,\bar{u}}$
by fitting the experimental data of $\phi$, $\Lambda$ and $\bar{\Lambda}$
at mid-rapidity ($\left|y\right|$<1.0) in Au+Au collisions. Second,
we use the obtained $v_{3,q}$ to calculate $v_{3}$ of other hadrons
$\Omega$, $\Xi$, $p$ and compare them with the experimental data.
The extracted $v_{3,q}$ at three collision energies are shown in
Fig.~\ref{fig:fig5}. Results of hadrons $\phi$, $\Lambda$, $\Xi^{-}$,
$\Omega^{-}$, $\bar{\Lambda}$, $\bar{\Xi}^{+}$, $\bar{\Omega}^{+}$,
$p$ and $\bar{p}$ in 10-40\% and 0-80\% centralities in Au+Au collisions
at $\sqrt{s_{NN}}=$19.6, 54.4 and 200 GeV are shown in Fig.~\ref{fig:fig6}$-$\ref{fig:fig8}.
Experimental data of the $v_{3}$ of these hadrons (symbols in Figs.~
\ref{fig:fig6}$-$\ref{fig:fig8}) are taken from Refs. \citep{Dixit:2022geb,STAR:2022ncy,STAR:2022tfp}.

The data points of $v_{3}$ of hadrons are relatively few in comparison
with those of $v_{2}$ of hadrons. By comparing with those limited
experimental data, we see that theoretical calculations of $v_{3}$
of $\Omega$, $\Xi$, $p$ are generally in agreement with the available
experimental data in (at) the studied collision centralities (energies). 

\section{THE QUADRANGULAR FLOW OF IDENTIFIED HADRONS\label{sec:v4}}

 In this section, we study the $v_{4}$ of identified hadrons $\phi$,
$p$, $\Lambda$, $\Xi$, $\Omega$ at mid-rapidity ($\left|y\right|$<1.0)
in Au+Au collisions at $\sqrt{s_{NN}}=$ 200 GeV, and make comparisons
with the available experimental data \citep{Sharma:2015gqe,Li:2010hea,PHENIX:2014uik}.
According to Eqs. (\ref{eq:vkm_final}) and (\ref{eq:vkb_final}),
$v_{4}$ flows of hadrons are 
\begin{align}
v_{4,M_{j}} & \approx v_{4,q_{1}}\left[1+\left(\frac{a_{6}}{a_{4}}-2\right)v_{2,q_{1}}v_{2,\bar{q}_{2}}\right]+v_{4,\bar{q}_{2}}\left[1+\left(\frac{a_{6}}{a_{4}}-2\right)v_{2,q_{1}}v_{2,\bar{q}_{2}}\right]+v_{2,q_{1}}v_{2,\bar{q}_{2}},\label{eq:v4_Mi}\\
v_{4,B_{j}} & \approx v_{4,q_{1}}\left[1+\left(\frac{a_{6}}{a_{4}}-2\right)\left(v_{2,q_{1}}v_{2,q_{2}}+v_{2,q_{1}}v_{2,q_{3}}\right)\right]+v_{4,q_{2}}\left[1+\left(\frac{a_{6}}{a_{4}}-2\right)\left(v_{2,q_{1}}v_{2,q_{2}}+v_{2,q_{2}}v_{2,q_{3}}\right)\right]\nonumber \\
 & +v_{4,q_{3}}\left[1+\left(\frac{a_{6}}{a_{4}}-2\right)\left(v_{2,q_{1}}v_{2,q_{3}}+v_{2,q_{2}}v_{2,q_{3}}\right)\right]+v_{2,q_{1}}v_{2,q_{2}}\left[1+\frac{1}{2}\left(\frac{v_{3,q_{2}}}{v_{2,q_{2}}}+\frac{v_{3,q_{1}}}{v_{2,q_{1}}}\right)v_{3,q_{3}}\right]\nonumber \\
 & +v_{2,q_{1}}v_{2,q_{3}}\left[1+\frac{1}{2}\left(\frac{v_{3,q_{1}}}{v_{2,q_{1}}}+\frac{v_{3,q_{3}}}{v_{2,q_{3}}}\right)v_{3,q_{2}}\right]+v_{2,q_{2}}v_{2,q_{3}}\left[1+\frac{1}{2}\left(\frac{v_{3,q_{2}}}{v_{2,q_{2}}}+\frac{v_{3,q_{3}}}{v_{2,q_{3}}}\right)v_{3,q_{1}}\right].\label{eq:v4_Bi}
\end{align}
We see that $v_{4}$ of hadrons mainly depends not only on the summation
of $v_{4}$ of constituent (anti-)quarks but also the product of the
lower-order flow $v_{2}$ of two (anti-)quarks. The modification factor
after $v_{4,q}$ deviates from one by the order of $v_{2,q}^{2}$
$\sim$$10^{-3}$. The modification factor after $v_{2,q_{1}}v_{2,q_{2}}$
deviates from one by the term $\left(v_{3,q_{1}}/v_{2,q_{1}}\right)v_{3,q_{2}}$
which is equal to $a_{3}^{2}v_{2,q_{1}}v_{2,q_{2}}$ according to
the scaling assumption Eq. (\ref{eq:vnq_scaling}) and therefore is
also very small. Therefore, we neglect these small modifications and
obtain 
\begin{align}
v_{4,M_{j}}\left(p_{T}\right) & \approx v_{4,q_{1}}\left(x_{1}p_{T}\right)+v_{4,\bar{q}_{2}}\left(x_{2}p_{T}\right)+v_{2,q_{1}}\left(x_{1}p_{T}\right)v_{2,\bar{q}_{2}}\left(x_{2}p_{T}\right),\label{eq:v4_Mi-4}\\
v_{4,B_{j}}\left(p_{T}\right) & \approx v_{4,q_{1}}\left(x_{1}p_{T}\right)+v_{4,q_{2}}\left(x_{2}p_{T}\right)+v_{4,q_{3}}\left(x_{3}p_{T}\right)+v_{2,q_{1}}\left(x_{1}p_{T}\right)v_{2,q_{2}}\left(x_{2}p_{T}\right)+v_{2,q_{1}}\left(x_{1}p_{T}\right)v_{2,q_{3}}\left(x_{3}p_{T}\right)\nonumber \\
 & +v_{2,q_{2}}\left(x_{2}p_{T}\right)v_{2,q_{3}}\left(x_{3}p_{T}\right),\label{eq:v4_Bi-4}
\end{align}
where $r=m_{s}/m_{u}$. 

Using Eqs. (\ref{eq:v4_Mi-4}$-$\ref{eq:v4_Bi-4}), we obtain $v_{4}$
of identified hadrons 
\begin{align}
v_{4,p}\left(p_{T}\right) & =3v_{4,u}\left(\frac{p_{T}}{3}\right)+3v_{2,u}^{2}\left(\frac{p_{T}}{3}\right),\label{eq:v4,p}\\
v_{4,\Omega}\left(p_{T}\right) & =3v_{4,s}\left(\frac{p_{T}}{3}\right)+3v_{2,s}^{2}\left(\frac{p_{T}}{3}\right),\label{eq:v4,omg}\\
v_{4,\phi}\left(p_{T}\right) & =2v_{4,s}\left(\frac{p_{T}}{2}\right)+v_{2,s}^{2}\left(\frac{p_{T}}{2}\right),\label{eq:v4,phi}\\
v_{4,\Lambda}\left(p_{T}\right) & =2v_{4,u}\bigl(\frac{1}{2+r}p_{T}\bigr)+v_{4,s}\bigl(\frac{r}{2+r}p_{T}\bigr)+2v_{2,u}\bigl(\frac{1}{2+r}p_{T}\bigr)v_{2,s}\bigl(\frac{r}{2+r}p_{T}\bigr)+v_{2,u}^{2}\bigl(\frac{1}{2+r}p_{T}\bigr),\label{eq:v4,lamb}\\
v_{4,\Xi}\left(p_{T}\right) & =2v_{4,s}\bigl(\frac{r}{1+2r}p_{T}\bigr)+v_{4,u}\bigl(\frac{1}{1+2r}p_{T}\bigr)+2v_{2,s}\bigl(\frac{r}{1+2r}p_{T}\bigr)v_{2,u}\Bigl(\frac{1}{1+2r}p_{T}\bigr)+v_{2,s}^{2}\bigl(\frac{r}{1+2r}p_{T}\bigr),\label{eq:v4,xi}
\end{align}
here, we have adopted the isospin symmetry between up and down quarks
and charge conjugation symmetry between strange quarks and strange
anti-quarks. 

\begin{figure}
\centering{}\includegraphics[width=0.6\linewidth]{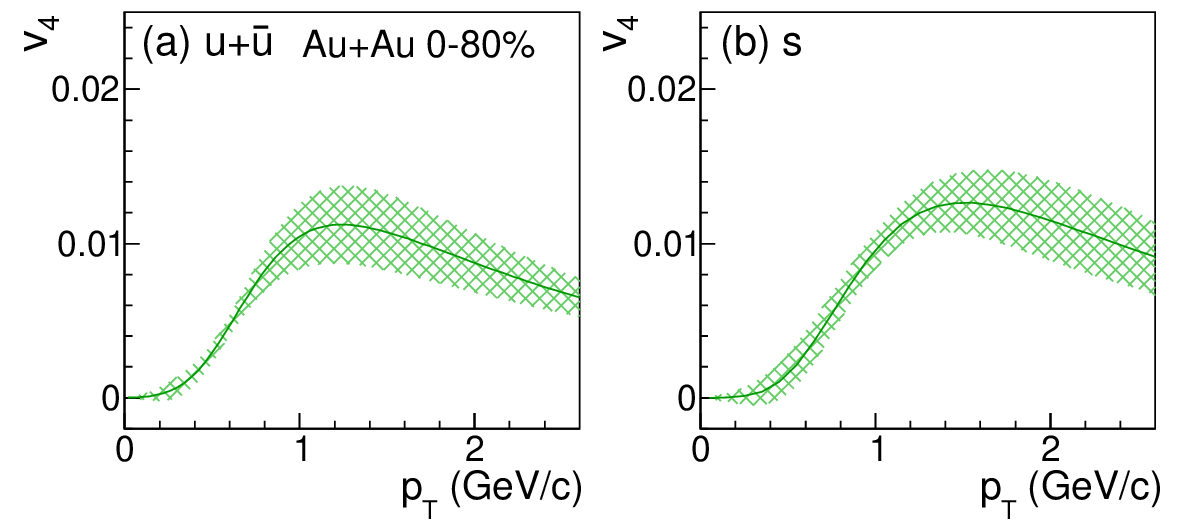}\caption{$v_{4}$ of quarks as a function of $p_{T}$ in 0-80\% centrality
in Au+Au collisions at $\sqrt{s_{NN}}=$ 200 GeV. \label{fig:fig9}}
\end{figure}

\begin{figure}
\includegraphics[width=0.65\linewidth]{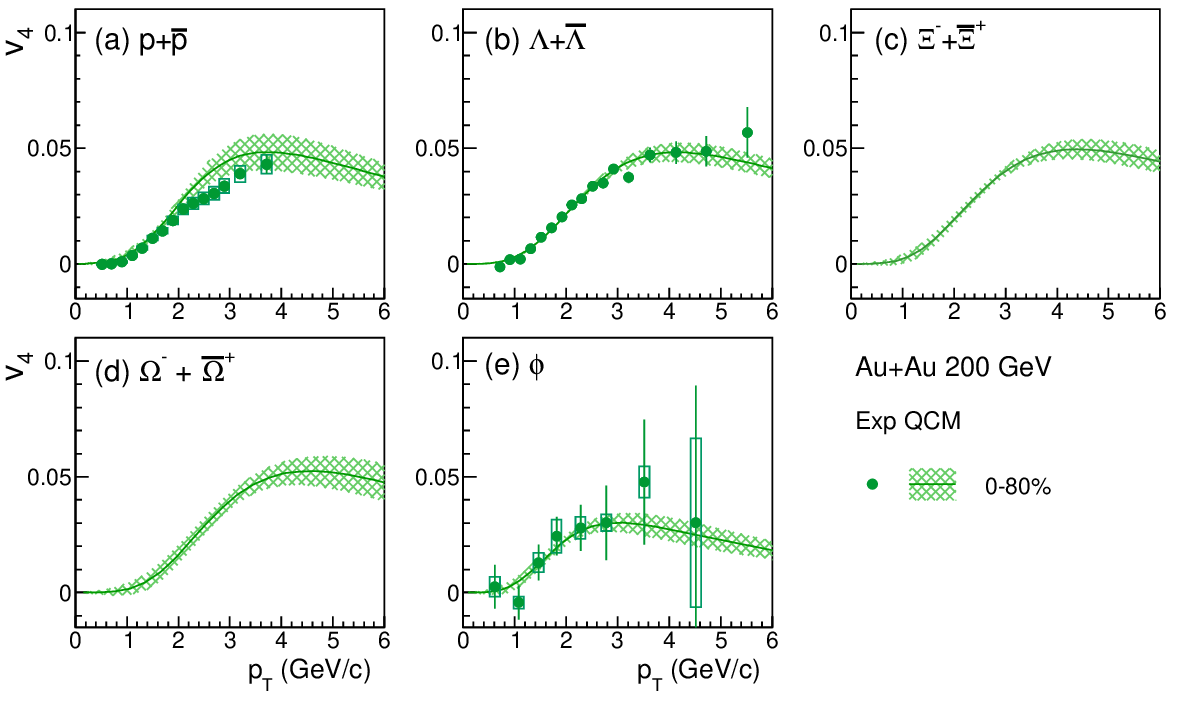}\caption{$v_{4}$ as a function of $p_{T}$ at mid-rapidity($\left|y\right|<1.0$)
for 0-80\% centrality in Au+Au collisions at $\sqrt{s_{NN}}=$200
GeV. Symbols are the experimental data \citep{Sharma:2015gqe,Li:2010hea,PHENIX:2014uik}
and lines are model results. \label{fig:fig10}}
\end{figure}

We apply the experimental data of identified hadrons $\phi$, $\Lambda$,
$\Xi$, $\Omega$ and $p$ for the 0-80\% centrality in Au+Au collisions
at $\sqrt{s_{NN}}=$ 200 GeV to test the above formulas. Following the
procedure in Sec. \ref{sec:v2} and Sec. \ref{sec:v3}, we first
apply the parameterized form in Eq. (\ref{eq:v2q}) to extract the
$v_{4,s}$ and $v_{4,u}$+$v_{4,\bar{u}}$ from the experimental data
of $\phi$ and $\Lambda$+$\bar{\Lambda}$ and use the extracted
quark $v_{4}$ to calculate $v_{4}$ of other hadrons. Fig.~\ref{fig:fig9}
shows the $v_{4}$ of quarks in 0-80\% centrality in Au+Au collisions
at $\sqrt{s_{NN}}=$ 200 GeV. Fig.~\ref{fig:fig10} shows theoretical
calculations of $\Xi$, $\Omega$ and $p$ and the comparison with
the available experimental data of $p+\bar{p}$. Due to the nonlinear
feature of the product term $v_{2,q_{1}}v_{2,q_{2}}$, the proportion
of the quark $v_{2}$ product term depends on $p_{T}$ and reaches
its maximum in the intermediate $p_{T}$ region. In general, the contribution
of $v_{2,q_{1}}v_{2,q_{2}}$ product terms to hadronic $v_{4}$ is
about 25\% and that of the summation of quark $v_{4}$ is about 75\%
in the intermediate $p_{T}$ region.

\section{The properties of $v_{2}$, $v_{3}$, and $v_{4}$ of quarks \label{sec:quark properties}}

\begin{figure}
\centering{}\includegraphics[width=0.65\linewidth,viewport=0bp 00bp 550bp 510bp]{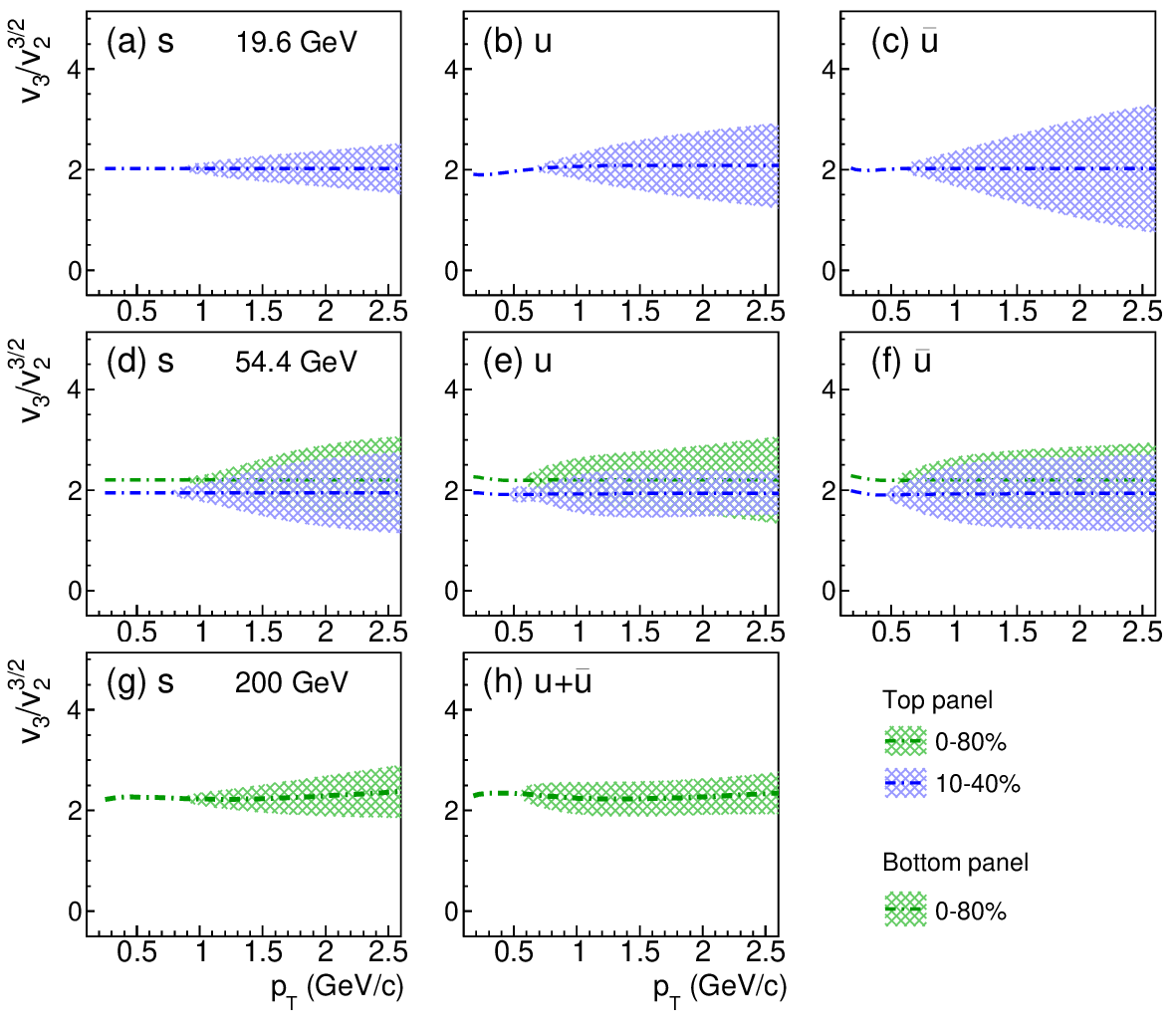}\caption{The ratio $v_{3}/v_{2}^{3/2}$ of quarks as a function of $p_{T}$
in different centralities in Au+Au collisions at $\sqrt{s_{NN}}=$
19.6, 54.4 and 200 GeV. \label{fig:fig11}}
\end{figure}

\begin{figure}
\centering{}\includegraphics[width=0.6\linewidth]{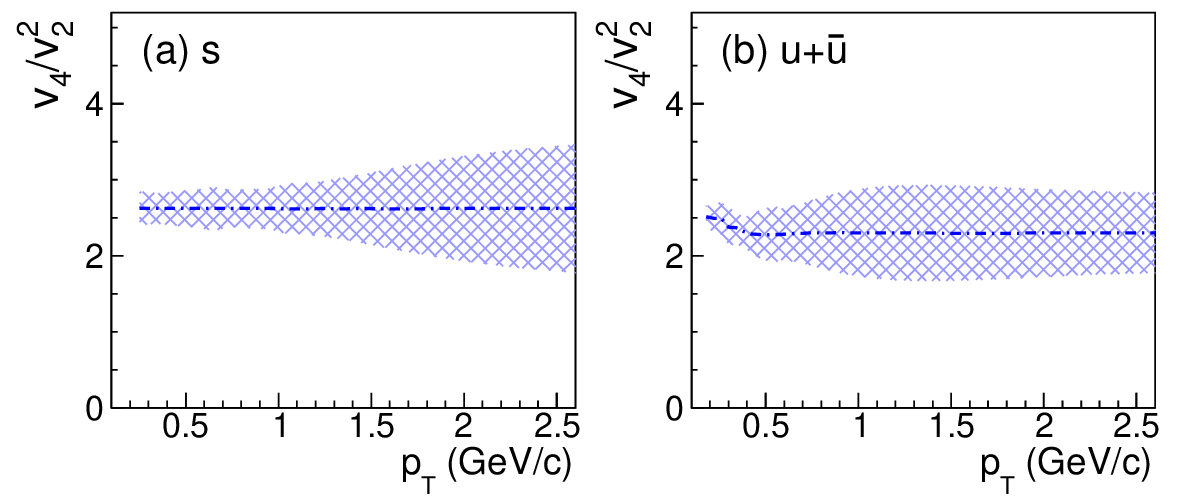}\caption{The ratio $v_{4}/v_{2}^{2}$ of quarks as a function of $p_{T}$ in
0-80\% centrality in Au+Au collisions at $\sqrt{s_{NN}}=$ 200 GeV. \label{fig:fig12}}
\end{figure}

In the above sections, we have obtained $v_{2}$, $v_{3}$ and $v_{4}$
of up/down quarks and strange quarks just before hadronization. In
this section, we study their properties by focusing on two possible
scaling relations. In Sec. \ref{sec:EVC_model}, we assume a scaling
relation $v_{n,q}=a_{n}v_{2,q}^{n/2}$ in Eq. (\ref{eq:vnq_scaling})
for quark $v_{n}$ in order to estimate the contribution of high order
flows of quarks to lower-order flows of hadrons. Here, we test this
relation by the obtained $v_{2}$, $v_{3}$, and $v_{4}$ of up/down
quarks and strange quarks. Results of $a_{3}=v_{3}/v_{2}^{3/2}$ and
$a_{4}=v_{4}/v_{2}^{2}$ as the functions of $p_{T}$ are shown in
Figs.~\ref{fig:fig11} and \ref{fig:fig12}. We see that $a_{3}$
and $a_{4}$ of up (anti-)quarks and those of strange quarks all exhibit
a weak dependence on $p_{T}$ and their values are both around two
and are less dependent on quark flavor. This verifies the consistency
of the scaling relation in Eq. (\ref{eq:vnq_scaling}) used in analyzing
$v_{n}$ flows of hadrons in current model. Above scaling property
of quark flows can be qualitatively understood by hydrodynamic evolution
of hot quark matter \citep{Borghini:2005kd,Retinskaya:2013gca}. 

The similar scaling relation should also exist at the hadron level. According to Eqs. (\ref{eq:vkm_final}) and (\ref{eq:vkb_final}) and discussions in the previous three sections, we have 
\begin{align}
v_{n,M_{j}}(p_{T}) & =v_{n,q_{1}}(x_{1}p_{T})+v_{n,\bar{q}_{2}}(x_{2}p_{T})+\sum_{k=2}^{n-2}v_{k,q_{1}}(x_{1}p_{T})v_{n-k,\overline{q}_{2}}(x_{2}p_{T}),\label{eq:vM-1}\\
v_{n,B_{j}}(p_{T}) & =v_{n,q_{1}}(x_{1}p_{T})+v_{n,q_{2}}(x_{2}p_{T})+v_{n,q_{3}}(x_{3}p_{T})\nonumber \\
 & +\sum_{k=2}^{n-2}\left[v_{k,q_{1}}(x_{1}p_{T})v_{n-k,q_{2}}(x_{2}p_{T})+v_{k,q_{2}}(x_{2}p_{T})v_{n-k,q_{3}}(x_{3}p_{T})+v_{k,q_{3}}(x_{3}p_{T})v_{n-k,q_{1}}(x_{1}p_{T})\right]\nonumber \\
 & +\sum_{k,m=2}^{m+k\leq n-2}v_{k,q_{1}}(x_{1}p_{T})v_{m,q_{2}}(x_{2}p_{T})v_{n-m-k,q_{3}}(x_{3}p_{T}),
\end{align}
after neglecting the effect of the modification terms in brackets
after $v_{n,q}$ in Eqs. (\ref{eq:vkm_final}) and (\ref{eq:vkb_final}).
Because $p_{T}/m_{M}=x_{1}p_{T}/m_{q_{1}}=x_{2}p_{T}/m_{q_{2}}$ in
EVC mechanism, the above formula can be expressed at the same transverse
velocity defined by $v_{\perp}=p_{T}/m_{M}$, 
\begin{align}
v_{n,M_{j}} & =v_{n,q_{1}}+v_{n,\bar{q}_{2}}+\sum_{k=2}^{n-2}v_{k,q_{1}}v_{n-k,\overline{q}_{2}},\label{eq:vM-1-1}\\
v_{n,B_{j}} & =v_{n,q_{1}}+v_{n,q_{2}}+v_{n,q_{3}}+\sum_{k=2}^{n-2}\left[v_{k,q_{1}}v_{n-k,q_{2}}+v_{k,q_{2}}v_{n-k,q_{3}}+v_{k,q_{3}}v_{n-k,q_{1}}\right]\nonumber \\
 & +\sum_{k,m=2}^{m+k\leq n-2}v_{k,q_{1}}v_{m,q_{2}}v_{n-m-k,q_{3}},
\end{align}
where we have hidden the variable $v_{\perp}$ in flow function of
quarks and those of hadrons. Using the scaling relation $v_{n,q}=a_{n}v_{2,q}^{n/2}$
in Eq. (\ref{eq:vnq_scaling}), we obtain 
\begin{align}
v_{n,M_{j}} & =a_{n}\left[v_{2,q_{1}}^{n/2}+v_{2,\bar{q}_{2}}^{n/2}\right]+\sum_{k=2}^{n-2}a_{k}v_{2,q_{1}}^{k/2}a_{n-k}v_{2,\bar{q}_{2}}^{(n-k)/2}\nonumber \\
 & \approx v_{2,q}^{n/2}\left[2a_{n}+\sum_{k=2}^{n-2}a_{k}a_{n-k}\right],\\
v_{n,B_{j}} & =a_{n}\left[v_{2,q_{1}}^{n/2}+v_{2,q_{2}}^{n/2}+v_{2,q_{3}}^{n/2}\right]+\sum_{k=2}^{n-2}a_{k}a_{n-k}\left[v_{2,q_{1}}^{k/2}v_{2,q_{2}}^{(n-k)/2}+v_{2,q_{3}}^{k/2}v_{2,q_{3}}^{(n-k)/2}+v_{2,q_{3}}^{k/2}v_{2,q_{1}}^{(n-k)/2}\right]\nonumber \\
 & +\sum_{k,m=2}^{m+k\leq n-2}a_{k}a_{m}a_{n-m-k}v_{2,q_{1}}^{k/2}v_{2,q_{2}}^{m/2}v_{2,q_{3}}^{(n-m-k)/2}\nonumber \\
 & \approx v_{2,q}^{n/2}\left[3a_{n}+3\sum_{k=2}^{n-2}a_{k}a_{n-k}+\sum_{k,m=2}^{m+k\leq n-2}a_{k}a_{m}a_{n-m-k}\right].
\end{align}
In the second line, we assume $v_{2,q_{1}}\approx v_{2,\bar{q}_{1}}$
and $v_{2,s}\approx v_{2,u}$ at the same transverse velocity in order to see a rough scaling relation among $v_{n,M_{j}}$
and $v_{n,B_{j}}.$ 

For low-order flows of hadrons, we obtain 
\begin{align}
v_{2,M_{j}} & \approx2v_{2,q},\\
v_{2,B_{j}} & \approx3v_{2,q},
\end{align}
and
\begin{align}
v_{3,M_{j}} & \approx2a_{3}v_{2,q}^{3/2}\approx\frac{a_{3}}{\sqrt{2}}v_{2,M_{j}}^{3/2},\\
v_{3,B_{j}} & \approx3a_{3}v_{2,q}^{3/2}\approx\frac{a_{3}}{\sqrt{3}}v_{2,B_{j}}^{3/2},
\end{align}
and
\begin{align}
v_{4,M_{j}} & \approx2\left(a_{4}+0.5\right)v_{2,q}^{2}\approx\frac{a_{4}+0.5}{2}v_{2,M_{j}}^{2},\\
v_{4,B_{j}} & \approx3\left(a_{4}+1\right)v_{2,q}^{2}\approx\frac{a_{4}+1}{3}v_{2,B_{j}}^{2},
\end{align}
at the same transverse velocity. We see that the $v_{3}$ and $v_{4}$
flows of hadrons also exhibit the similar scaling property $v_{n,h}\propto v_{2,h}^{n/2}$.
The ratio of $v_{3,M_{j}}/v_{2,M_{j}}^{3/2}$ is $\sqrt{3/2}-1\approx20\%$
greater than that of $v_{3,B_{j}}/v_{2,B_{j}}^{3/2}$ and the ratio
of $v_{4,M_{j}}/v_{2,M_{j}}^{2}$ is $(2a_{4}-1)/4(a_{4}+1)\approx25\%$
greater than that of $v_{4,B_{j}}/v_{2,B_{j}}^{2}$. These two properties
are qualitatively indicated by the measurement of higher order flows
of $\phi$, proton and hyperons in Au+Au collisions at RHIC energies
\citep{STAR:2021twy,Dixit:2022geb}. 

\begin{figure}[tbh]
\centering{}\includegraphics[width=0.65\linewidth,viewport=0bp 00bp 550bp 510bp]{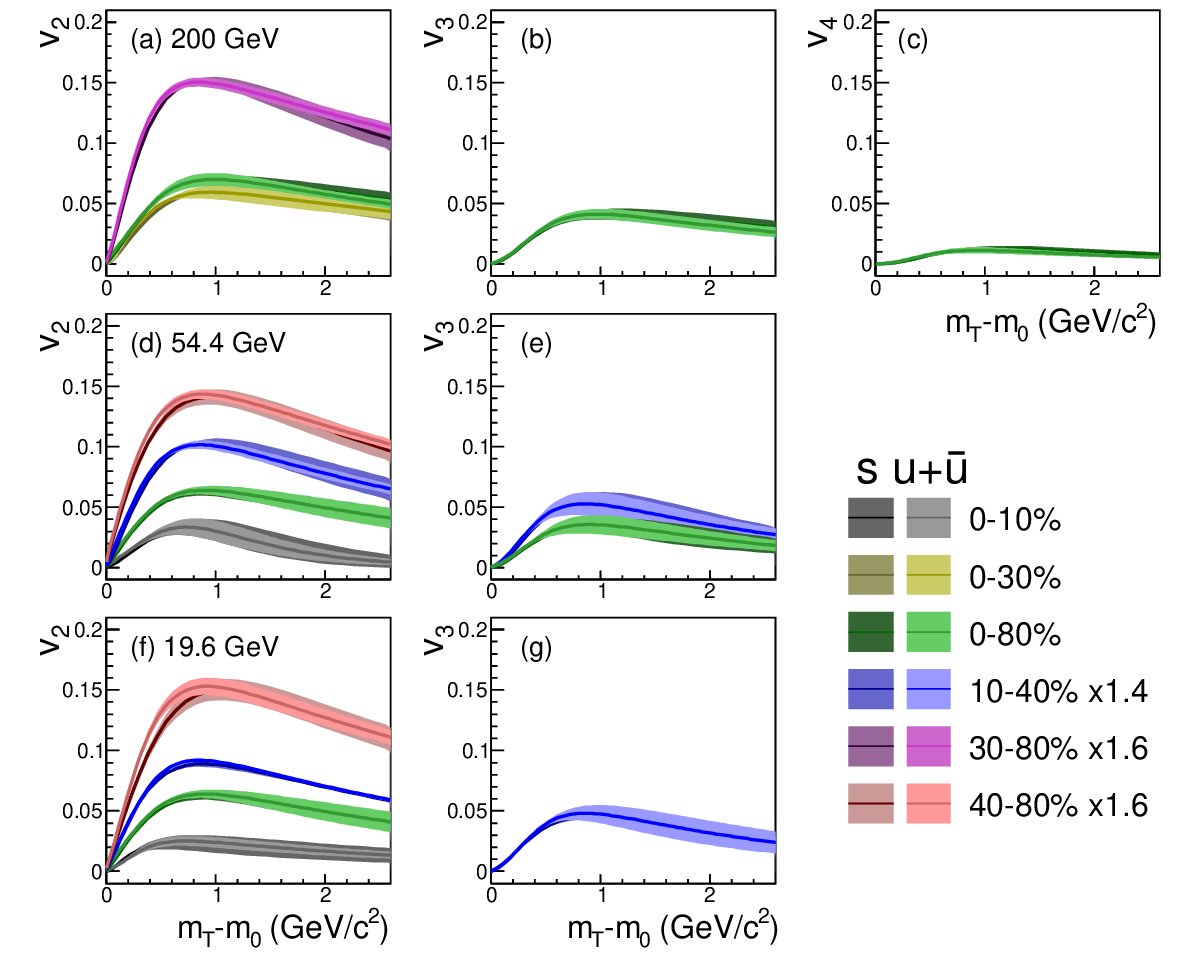}\caption{$v_{2}$, $v_{3}$ and $v_{4}$ of quarks as a function of the transverse
kinetic energy in different centralities in Au+Au collisions at $\sqrt{s_{NN}}=$19.6,
54.4 and 200 GeV. \label{fig:fig13}}
\end{figure}

Comparing the obtained $v_{2}$, $v_{3}$ and $v_{4}$ of up/down
quarks and strange quarks as functions of $p_{T}$ in Figs.~\ref{fig:fig1},
\ref{fig:fig5} and \ref{fig:fig9}, we can find a certain difference
between flow of up quarks and that of strange quarks. Now, we change
the kinetic variable $p_{T}$ to transverse kinetic energy $m_{T}-m_{0}$ ($KE_{T}$)
and show flows of strange quarks and up anti-quarks as the function
of $m_{T}-m_{0}$ in Fig.~\ref{fig:fig13}. In this case, we see that the difference between flows of strange quarks and those of up anti-quarks is small, which means a rough $KE_{T}$ scaling property for flows of quarks.  This property can be qualitatively understood to be the hydrodynamic evolution of anisotropic flows at quark level \citep{Csanad:2005gv}.

\section{Summary\label{sec:Summary}}

In this paper, we have applied the equal-velocity quark combination
model to study $v_{2}$, $v_{3}$ and $v_{4}$ of identified hadrons
$\phi$, $\Lambda$, $\Xi^{-}$, $\Omega^{-}$, $\bar{\Lambda}$,
$\bar{\Xi}^{+}$, $\bar{\Omega}^{+}$, $p$ and $\bar{p}$ as a function
of $p_{T}$ at mid-rapidity ($\left|y\right|<1.0$) for different
centralities (0-10\%, 0-80\%, 0-30\%, 10-40\%, 40-80\% and 30-80\%)
in Au+Au collisions at $\sqrt{s_{NN}}=$19.6, 54.4 and 200 GeV. We
derived formulas of anisotropic flows of hadrons under equal-velocity
quark combination mechanism and simplified $v_{2}$, $v_{3}$ and
$v_{4}$ of hadrons by neglecting contributions of higher order flows
of quarks at hadronization. Using these simplified formulas of hadronic
flows, we gave a self-consistent explanation on $v_{2}$, $v_{3}$
and $v_{4}$ of different hadrons measured by the STAR collaboration
in Au+Au collisions at RHIC BES energies. 

In the equal-velocity quark combination mechanism, flows of hadrons
are the simple summation of that of constituent quarks at the transverse
velocity the same as that of the formed hadron. This is different
from that in NCQ scaling where quarks have one half/third of the transverse
momentum or kinetic energy of the formed hadron. We used the experimental
data of $\Lambda$, $\bar{\Lambda}$ and $\phi$ to extract flows
of up quark, up anti-quark and strange quark just before hadronization.
Then we apply the obtained quark flows to calculate flows of other
hadrons such as $\Xi^{-}$, $\Omega^{-}$, $\bar{\Xi}^{+}$, $\bar{\Omega}^{+}$,
$p$, $\bar{p}$ and compare with experimental data of these hadrons.
The systematical comparison between our theoretical prediction of
$v_{2}$ and $v_{3}$ of hadrons with the experimental data in Au+Au
collisions at $\sqrt{s_{NN}}=$19.6, 54.4 and 200 GeV suggest a good
self-consistency of equal-velocity quark combination mechanism. The
model description in $v_{4}$ of hadrons in Au+Au collisions at $\sqrt{s_{NN}}=$200
GeV is also in good agreement with the available experimental data
of the STAR collaboration. 

We also studied the properties of the obtained $v_{2}$, $v_{3}$
and $v_{4}$ of constituent quarks just before hadronization in Au+Au
collisions at the studied collision energies and found two possible
properties. One is the scaling property in different order flows of
quarks, i.e., $v_{n,q}=a_{n}v_{2,q}^{n/2}$, where $a_{3}$ and $a_{4}$
are about two and weakly depend on the transverse momentum, collision
energy and collision centrality. Another is the approximate scaling
for flows of up/down quarks and strange quarks, $v_{2,u}\approx v_{2,s}$
at the same transverse kinetic energy $m_{T}-m_{0}$. These two properties
can be understood if we follow the hydrodynamic evolution of QGP and
apply the thermal freeze-out scenario to the momentum distribution
of constituent quarks.

\section{Acknowledgments}

This work was supported in part by the National Natural Science Foundation
of China under Grant No. 12375074 and 12175115, and Higher Educational
Youth Innovation Science and Technology Program of Shandong Province
under Grants No. 2020KJJ004.

\bibliographystyle{apsrev4-1}
\bibliography{ref}

\end{document}